\documentclass[trackchanges,twocolumn]{aastex701}
\usepackage{multirow}
\usepackage[figuresright]{rotating}
\usepackage{subfigure}
\usepackage{epic,eepic}
\usepackage{graphicx}
\usepackage{longtable}
\usepackage{float}
\usepackage{lineno}
\usepackage{pifont}
\usepackage{gensymb}
\usepackage{booktabs}
\usepackage{graphicx}
\usepackage{threeparttable}
\usepackage{textcomp, gensymb}
\usepackage{amsmath}
\usepackage{color}
\usepackage{CJK}
\usepackage{hyperref}
\usepackage{soul}

\shorttitle{\rm X-ray activity evolution}
\shortauthors{Li et al.}

\begin{document}
\begin{CJK*}{UTF8}{gbsn}


\title{\rm Evolution of Stellar Activity and Habitable Zone (EATEN): III. X-ray Activity of Dwarfs in Open Clusters and Field Stars}


\author[orcid=0009-0007-4501-4376]{Xue Li}
\affiliation{School of Astronomy and Space Science, University of Chinese Academy of Sciences, Beijing 100049, People's Republic of China}
\email{lixue@bao.ac.cn}

\author[orcid=0000-0003-3116-5038]{Song Wang}
\affiliation{National Astronomical Observatories, Chinese Academy of Sciences, Beijing 100101, People's Republic of China}
\affiliation{Institute for Frontiers in Astronomy and Astrophysics, Beijing Normal University, Beijing 102206, People's Republic of China}
\email[show]{songw@bao.ac.cn}

\author[orcid=0000-0003-3474-5118]{Henggeng Han}
\affiliation{National Astronomical Observatories, Chinese Academy of Sciences, Beijing 100101, People's Republic of China}
\email{hghan@nao.cas.cn}

\author[orcid=0000-0001-6329-6644]{Jun Ma}
\affiliation{School of Astronomy and Space Science, University of Chinese Academy of Sciences, Beijing 100049, People's Republic of China}
\email{majun@nao.cas.cn}

\author[orcid=0000-0003-3250-2876]{Yang Huang}
\affiliation{School of Astronomy and Space Science, University of Chinese Academy of Sciences, Beijing 100049, People's Republic of China}
\affiliation{National Astronomical Observatories, Chinese Academy of Sciences, Beijing 100101, People's Republic of China}
\email{huangyang@ucas.ac.cn}

\author[orcid=0000-0002-2874-2706,gname=Jifeng, sname='Liu']{Jifeng Liu} 
\affiliation{School of Astronomy and Space Science, University of Chinese Academy of Sciences, Beijing 100049, People's Republic of China}
\affiliation{New Cornerstone Science Laboratory, National Astronomical Observatories, Chinese Academy of Sciences, Beijing 100012, People's Republic of China}
\email{jfliu@nac.cas.cn}

\let\cleardoublepage\clearpage
\begin{abstract}
Stellar X-ray emission serves as a direct diagnostic of coronal activity, which is fundamentally linked to coronal heating processes. It also strongly influences the atmospheres and long-term habitability of orbiting exoplanets. Investigating how this high-energy emission evolves is therefore essential for understanding the evolution of stellar magnetic dynamos and planetary atmospheres and habitability.
In this work, we investigate the evolution of X-ray activity and XUV irradiation for a sample of F-M dwarf stars based on \textit{Chandra} and \textit{XMM-Newton} observations. 
We find that F- and G-type stars broadly follow the traditional evolutionary picture of an early saturated (or weakly declining) phase followed by a modest decline, whereas K- and M-type stars exhibit a clear three-phase evolution of a saturated phase, an intermediate phase of rapid decay, and a final modest decline phase.
By combining X-ray, ultraviolet, and Ca~II~H\&K bands, we show that coronal emission becomes increasingly dominant toward lower-mass stars. 
Based on the cumulative XUV emission calculated from our fitted relation, planets around F- and G-type stars experience relatively moderate XUV environments, while those around K- and M-type stars may exceed the empirical cosmic shoreline shortly after reaching the main sequence, though this conclusion depends on the adopted shoreline value.

\end{abstract}


\section{INTRODUCTION}
\label{intro.sec}

Magnetic activity is an intrinsic property of stars and arises from dynamo processes that are related to stellar rotation. Young, rapidly rotating stars typically generate strong magnetic fields and exhibit high levels of chromospheric and coronal activity, whereas older stars undergo rotational spin-down driven by magnetic braking and display progressively declining activity levels \citep[e.g.,][]{1972ApJ...171..565S,2012MNRAS.422.2024J,2013MNRAS.431.2063S}. 

As stars evolve on the main sequence, angular momentum loss through magnetized winds leads to a systematic decrease in rotation rate, resulting in a weakening of dynamo efficiency and magnetic activity \citep[e.g.,][]{2008ApJ...687.1264M,2011ApJ...743...48W, 2014ApJ...794..144R}. Among activity diagnostics, X-ray emission traces million-degree coronal plasma heated by magnetic reconnection and wave dissipation, and thus serves as a particularly sensitive probe of stellar magnetic evolution.

Previous studies have established a broad evolutionary picture in which young, rapidly rotating stars occupy a saturated regime of X-ray emission, characterized by an approximately constant ratio of X-ray to bolometric luminosity, followed by a decline in X-ray activity as stellar rotation slows on the main sequence \citep[e.g.,][]{2012MNRAS.422.2024J,2016ApJ...830...44N,2017MNRAS.471.1012B}. However, reported decay slopes and transition ages vary substantially among different studies. These discrepancies may be caused by heterogeneous sample selections, limited age coverage, particularly the lack of stars with intermediate and old ages ($\gtrsim 1$~Gyr), and the mixing of samples with different spectral types and metallicities \citep[e.g.,][]{2012MNRAS.422.2024J,2016ApJ...830...44N}.

This phenomenon becomes increasingly notable toward lower stellar masses. For K- and M-type main-sequence stars, reliable age estimates are particularly difficult to obtain due to their long evolutionary timescales. As a result, only a few data points are available in the old-age regime of the X-ray activity--age relations for these stars, which can easily lead to different relation morphologies.

High-energy stellar radiation plays a crucial role in shaping the atmospheres and long-term habitability of orbiting exoplanets. X-ray and extreme-ultraviolet photons can drive upper-atmospheric heating and hydrodynamic escape, potentially leading to substantial atmospheric erosion \citep[e.g.,][]{2007AsBio...7..185L,2009A&A...506..399L,2013ApJ...775..105O,2019AREPS..47...67O}. Simultaneously, high-energy radiation regulates key photochemical and ionization processes that influence atmospheric structure composition \citep[e.g.,][]{2016AsBio..16...68R,2017ApJ...843..110R}. Quantifying how stellar high-energy emission evolves with age is therefore essential for understanding atmospheric loss processes, and for evaluating the long-term evolution of planetary atmospheres and habitable zones. 

In this work, we construct a large sample of dwarf stars spanning spectral types F--M, drawn from both open clusters and the field. We cross-matched them with X-ray observations from \textit{Chandra} and \textit{XMM--Newton}. Together with well-characterized stellar ages, we aim to re-examine the long-term evolution of coronal activity across different spectral types and to provide constraints on the high-energy radiation environment of these stars.

This paper is part of a series investigating the evolution of stellar activity and its implications for habitable zones. In \citet{2025ApJS..281...13L} (hereafter Paper~I), we focused on ultraviolet activity indicators and established a unified evolutionary framework for stellar UV emission from F- to M-type stars. 
In \citet{2026arXiv260117715H} (hereafter Paper~II), we focused on activity evolutionary in Ca~II~H\&K band from G- to M-type stars.
Here, we extend this investigation to the X-ray regime. Section~\ref{sample.sec} describes the construction of the stellar sample and the X-ray data for open cluster members and field stars. In Section~\ref{evo.sec}, we present the evolution of X-ray activity for different types of stars and comparison with previous studies. In Section~\ref{disscuss.sec}, we compare X-ray evolution with ultraviolet and Ca~II~H\&K activity indicators, and explores the implications of high-energy stellar radiation for planetary habitability.

\section{Sample and Data Reduction}
\label{sample.sec}
The original sample is from Paper~I, including open cluster (OC) members and field stars. The OC members were compiled from multiple \textit{Gaia}-based catalogs \citep[e.g.,][]{2020A&A...640A...1C, 2024A&A...686A..42H, 2020AJ....160..279K}, while the field star catalog with age measurements was taken from from \citet{2025ApJS..280...13W}, with stars selected to have spectral signal-to-noise ratios (S$/$Ns) of S/N$_g > 10$ and S/N$_i > 10$. We cross-matched this sample with the \textit{Chandra} and \textit{XMM-Newton} source catalogs to identify stars with available X-ray observations.

For OC members, extinction values were adopted from the cluster catalogs (see Paper~I for details), while those for field stars were estimated using the \texttt{mwdust} package based on their Galactic coordinates and distances \citep{2016ApJ...818..130B}, assuming $R_V=3.1$. We only retained dwarf stars with extinction ($E(B-V) < 0.5$) and distances smaller than 5 kpc. 

\subsection{\textit{Chandra}}
\label{chandra_data.sec}
The \textit{Chandra} X-ray Observatory \citep{2002PASP..114....1W} is one of NASA’s Great Observatories and provides the highest spatial resolution currently achievable in X-ray astronomy. The on-axis resolution is about 0.5\arcsec, which enables precise source localization. \textit{Chandra} operates over the $0.1-10$~keV energy range and is equipped with the Advanced CCD Imaging Spectrometer (ACIS; \citealt{2003SPIE.4851...28G}), which provides sensitive imaging and moderate spectral resolution. In this work, we make use of the second major release of the \textit{Chandra} Source Catalog (CSC~2.1; \citealt{2010ApJS..189...37E, 2024ApJS..274...22E}), which incorporates observations from 1999 to 2021 and contains 407,806 unique X-ray sources. This paper employs a list of \textit{Chandra} datasets, contained in the \textit{Chandra} Data Collection ~\dataset[DOI: 10.25574/cdc.623]{https://doi.org/10.25574/cdc.623}. 

We cross-matched our stellar sample with CSC~2.1 using a matching radius of 5\arcsec. We applied several criteria to ensure reliable X-ray detections, requiring \texttt{sat\_src\_flag = FALSE} to exclude saturated sources and \texttt{streak\_src\_flag = FALSE} to remove detections affected by readout streaks. In addition, we selected sources detected at a significance level of at least $2\sigma$ in the broad ($0.5-7$~keV) band with \texttt{flux\_significances\_b}~$\geq 2$. 
During the cross-matching process, we found that some \textit{Chandra} sources may be associated with multiple \textit{Gaia} counterparts, likely due to the broadening of the point-spread function at large off-axis angles. Therefore, if the matched \textit{Gaia} sources differ in $G$-band magnitude by more than 2.5~mag, we adopt the brightest source as the counterpart; otherwise, we select the source with the smallest angular separation from the \textit{Chandra} position. We obtained 878 unique sources with a total of 3,342 individual exposures, including 729 OC members and 129 field stars.

\subsection{\textit{XMM-Newton}}
\label{xmm_data.sec}
The \textit{XMM-Newton} Observatory is a European Space Agency X-ray mission with a large light-collecting area \citep{2001A&A...365L...1J}. 
It operates over the energy range from 0.2~keV to 12~keV and carries three co-aligned X-ray telescopes, each equipped with European Photon Imaging Cameras (EPIC), consisting of two MOS CCD detectors and one pn CCD detector. 
The latest \textit{XMM-Newton} Serendipitous Source Catalog (4XMM-DR14; \citet{2020A&A...641A.136W}) contains 1,035,832 detections corresponding to 692,109 unique X-ray sources\footnote{\url{http://xmmssc.irap.omp.eu/Catalogue/4XMM-DR14/4XMM_DR14.html}}.

We also cross-matched our sample with the 4XMM–DR14 catalog using a radius of 5\arcsec. Then we applied additional selections to ensure reliable detections in the EPIC cameras. We used only the EPIC-pn measurements in this work. We required \texttt{obs\_class}~$\leq 3$, excluding observations affected by high background levels, in particular soft-proton flares, or other data-quality issues that compromise reliable X-ray photometry \citep{2020A&A...641A.136W}.
We required \texttt{pn\_filter}~$\neq$~\texttt{UNDEF}, ensuring that the EPIC-pn filter configuration is properly defined. Sources with \texttt{sum\_flag = 0} were retained, corresponding to detections without any quality warnings in the 4XMM catalog. 
Finally, we imposed \texttt{pn\_8\_rate/pn\_8\_rate\_err}~$\geq 2$, requiring at least a $2\sigma$ detection in the EPIC-pn $0.2-12$~keV band \footnote{\url{https://heasarc.gsfc.nasa.gov/W3Browse/xmm-newton/xmmssc.html}}. In cases where a single XMM source matched multiple \textit{Gaia} counterparts, we applied the same procedure as for \textit{Chandra}. We then obtained 3,239 unique OC members with 7,046 individual exposures and 3460 unique field stars with 6,517 individual exposures.

\subsection{Sample Cleaning}
\label{sample_selection.sec}

When both \textit{Chandra} and \textit{XMM-Newton} measurements are available, we adopted the \textit{Chandra} result because its superior spatial resolution reduces potential source confusion. However, we also understand that when no additional \textit{Chandra} sources are detected within  $\approx$15'', \textit{XMM-Newton} may provide a more precise flux measurement due to its larger effective area. 

We applied dwarf selection defined in Paper~I to both the OC and the field stars, to identify stars located within the fitted dwarf area. In addition, we performed several procedures to exclude possible contaminants: (1) We cross-matched these sources with the SIMBAD database within 10\arcsec\ to identify binaries, Galaxies, or AGNs. (2) We cross-matched our sample with variability catalogs from Kepler \citep{2013MNRAS.432.1203M, 2014ApJS..211...24M}, K2 \citep{2019ApJS..244...21S, 2020A&A...635A..43R}, ZTF \citep{2020ApJS..249...18C}, TESS \citep{2023ApJS..268....4F}, and ASAS-SN \citep{2023MNRAS.519.5271C} to check photometric binaries. 

\begin{figure*}
    \centering
    \subfigure[]{
    \includegraphics[width=0.46\textwidth]{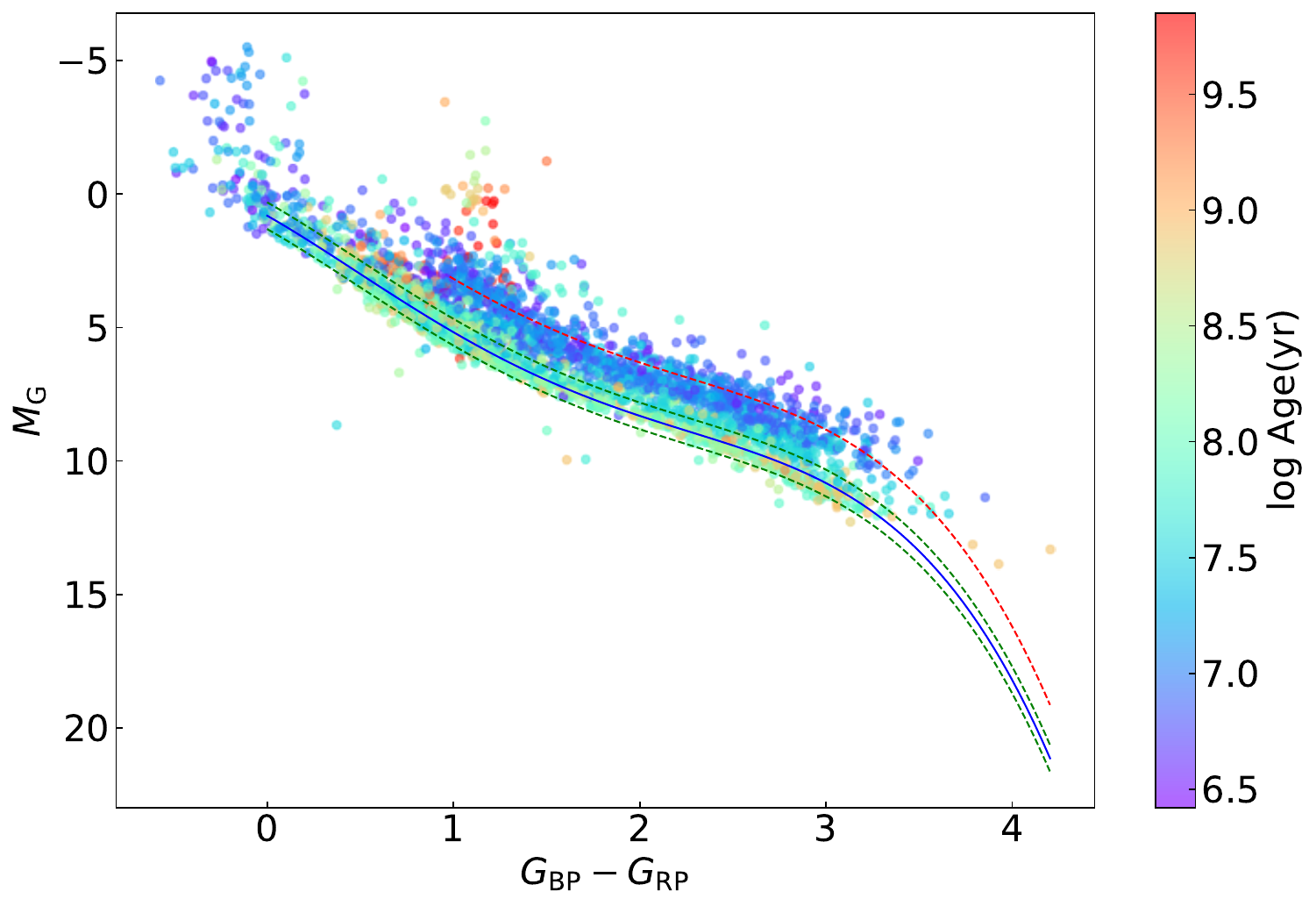}
    \label{hr_oc.fig}}
    \subfigure[]{
    \includegraphics[width=0.46\textwidth]{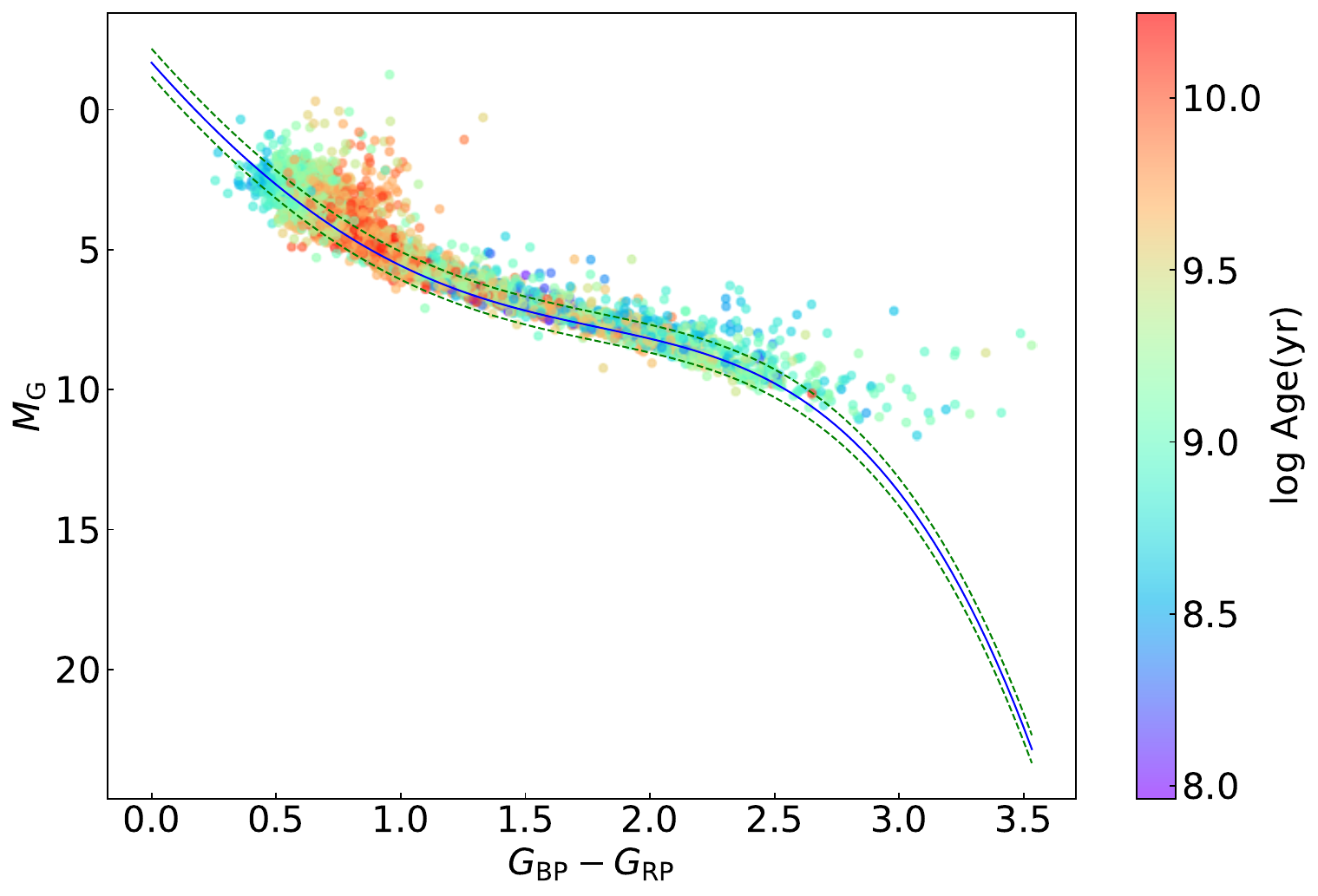}
    \label{hr_field.fig}}
    \subfigure[]{
    \includegraphics[width=0.46\textwidth]{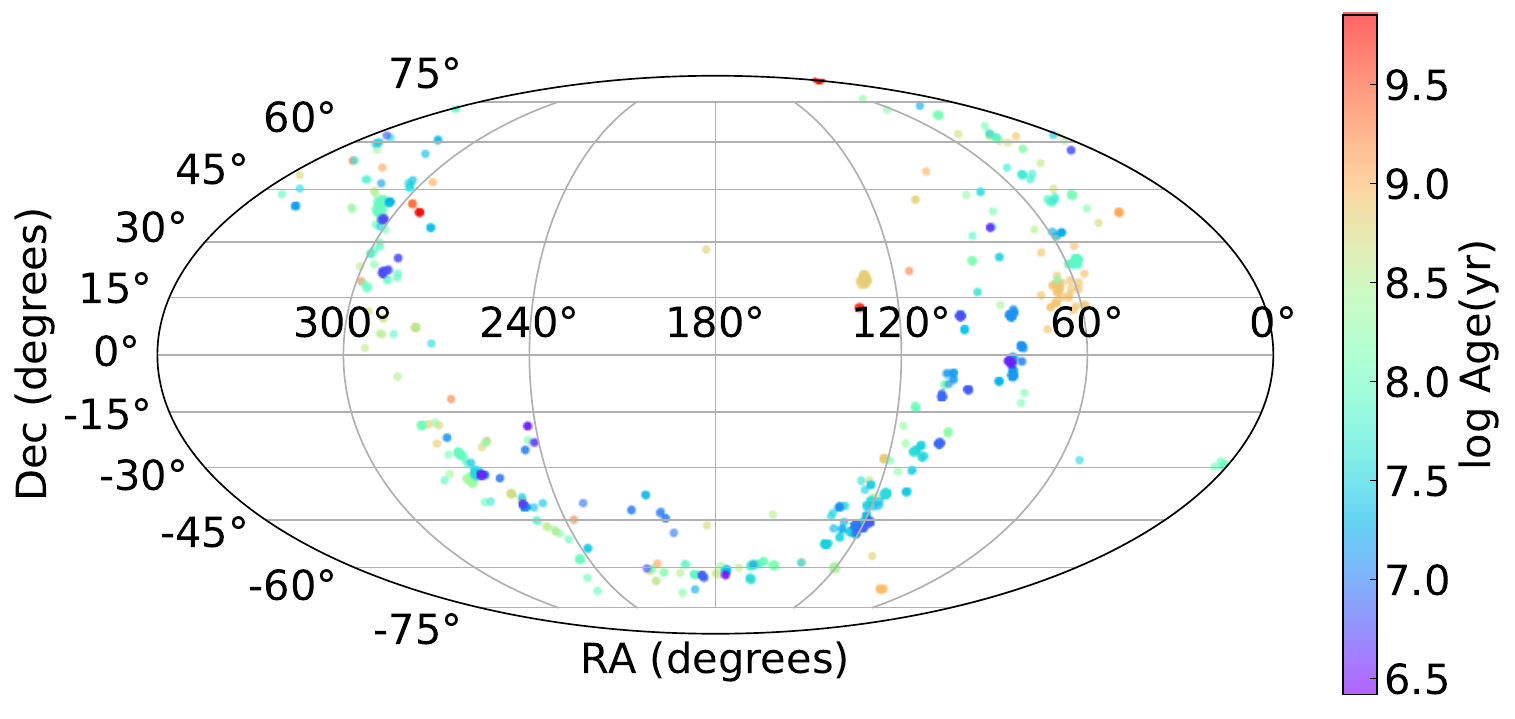}
    \label{galactic_oc.fig}}
    \subfigure[]{
    \includegraphics[width=0.46\textwidth]{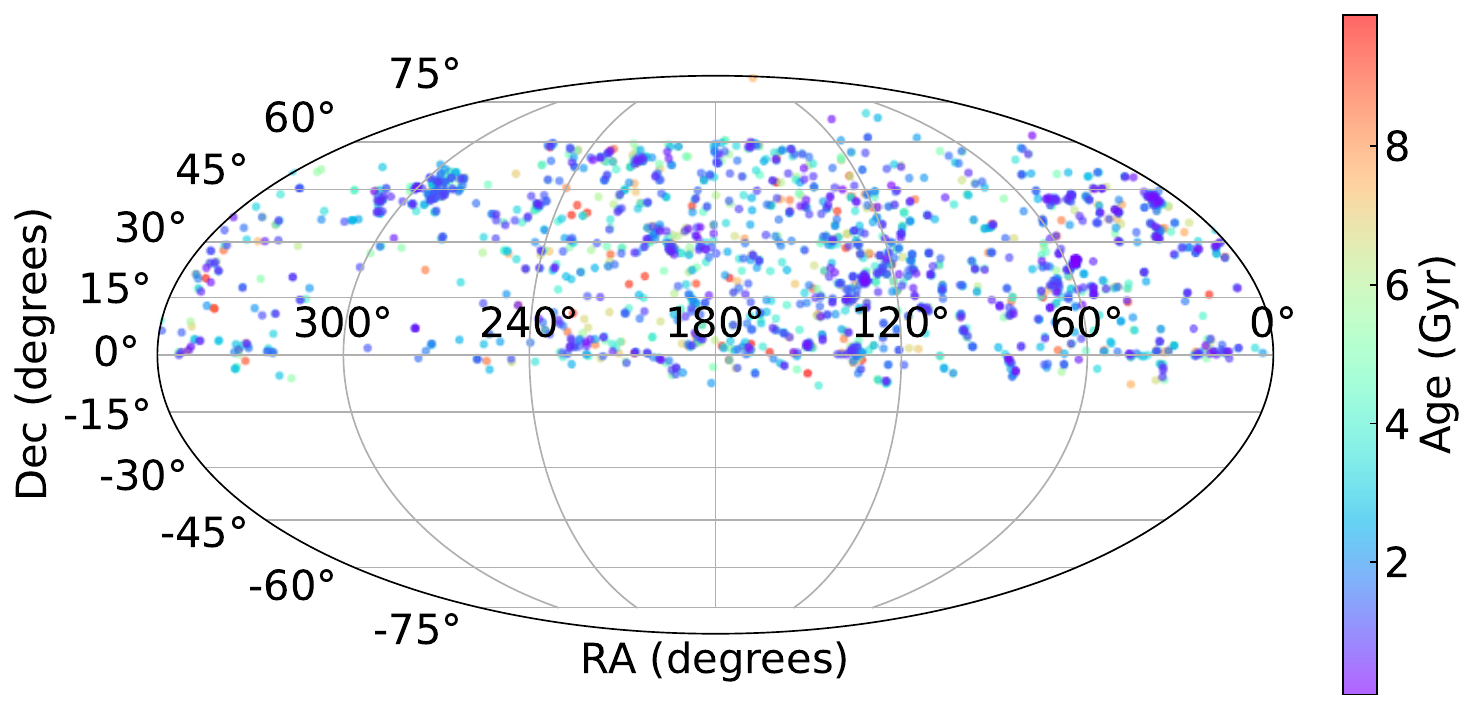}
    \label{galactic_field.fig}}
    \caption{Panel (a): H–R diagram of the OC members with detections. The blue solid line is the fitting results of MS. green dashed lines represent the upper and lower bounds of the MS fitted with a $\pm 0.5$ offset, and the red dashed line is the same sequence shifted upwards by 2. The color bars represent the stellar ages. 
    Panel (b): H–R diagram of the field stars with detections. 
    Panel (c): The Galactic distribution of the selected sources from the detected OC members. 
    Panel (d): The Galactic distribution of the selected sources from the detected field stars.}
    \label{hr.fig}
\end{figure*}

Our final dwarf catalog contains 2,958 stars, including 150 \textit{Chandra} sources (118 OC members and 32 field stars) and 2,808 \textit{XMM-Newton} sources (1,734 OC members and 1,074 field stars). 
The resulting Hertzsprung-Russell diagrams for the OC and field samples are shown in Figure~\ref{hr_oc.fig} and \ref{hr_field.fig}.
For the final set of selected dwarf stars, we also examined their Galactic spatial distribution, as illustrated in Figure~\ref{galactic_oc.fig} and \ref{galactic_field.fig}.

\subsection{X-ray Luminosity Calculation}
\label{cal.sec}

We recomputed the X-ray fluxes using \texttt{PIMMS}\footnote{\url{https://heasarc.gsfc.nasa.gov/docs/software/tools/pimms.html}} in order to obtain consistent luminosities of the sample. \texttt{PIMMS} is a widely used tool that simulates instrumental responses across different X-ray missions and computes fluxes based on specific spectral models \citep{1993Legac...3...21M}. For each \textit{Chandra} source, we extracted the count rates in the ultra-soft ($u$; $0.2-0.5$~keV) and broad ($b$; $0.5-7.0$~keV) bands from the Chandra Source Catalog, and summed them to obtain the total $0.2-7.0$~keV count rate used as input to \texttt{PIMMS}. We adopted an optically derived extinction value $A_V$ for each star and converted it to the hydrogen column density using $N_{\rm H} = 2.19\times10^{21}\times A_V \;{\rm cm^{-2}}$ \citep{2017MNRAS.471.3494Z}.
The flux conversion is assumes an Astrophysical Plasma Emission Code (APEC) thermal plasma model with a fixed temperature of $\log T = 6.5$ and metallicity of $Z = 0.5\,Z_\odot$, following \citet{2020ApJ...902..114W}. 
All fluxes were computed as unabsorbed $0.1-2.4$~keV energy fluxes using the instrument configuration corresponding to each observation (e.g., ACIS-I or ACIS-S). 

For each \textit{XMM-Newton} source, we also supplied \texttt{PIMMS} with the EPIC-pn count rate in the $0.2-12$~keV band and adopted the corresponding optical blocking filter (Thin/Medium/Thick) reported for the observation. As in the \textit{Chandra} calculations, we assumed the same APEC plasma model and used \texttt{PIMMS} to compute the unabsorbed flux in the 0.1--2.4 keV band. 

Although stellar coronae are intrinsically multi-temperature and exhibit a range of metallicities, we adopted a single-temperature, single-metallicity APEC model ($\log T=6.5$, $Z=0.5\,Z_\odot$) to ensure a homogeneous count-rate-to-flux conversion. To evaluate the associated uncertainty, we repeated the conversion using a range of coronal temperatures ($\log T=6.3$--$6.7$) and metallicities ($Z=0.3$--$0.7\,Z_\odot$). The resulting variations in $L_{\rm X}$ and $R_{\rm X}$ are very small, indicating the temperature and metallicity dependence can be neglected.

Then, we computed the X-ray luminosity for each star using the standard relation: $L_{\rm X} = 4 \pi d^2 F_{\rm X}$, where $F_{\rm X}$ is the unabsorbed flux in the $0.1-2.4$~keV band, and $d$ is the stellar distance obtained from OC catalogs for OC members and derived from \textit{Gaia} DR3 parallaxes for field stars. For stars covered by multiple observations, the flux were combined using the exposure times as weights,
\begin{equation}
F_{\rm X} =
\frac{\sum_i F_{{\rm X},i}, t_{{\rm exp},i}}
{\sum_i t_{{\rm exp},i}},
\end{equation}
where $F_{{\rm X},i}$ and $t_{{\rm exp},i}$ are the flux and exposure time of observation $i$, respectively.

To estimate the bolometric luminosities ($L_{\rm bol}$) of our samples, we employed the PARSEC stellar evolution models \citep{2012MNRAS.427..127B}. We downloaded PARSEC isochrones computed for solar metallicity ([Fe/H] = 0) with fine age spacing ($\Delta \log {\rm age} = 0.02$). We also downloaded \textit{Gaia} magnitudes and corrected their extinction.
For each star, we selected model points satisfying the following conditions:
\begin{align}
&|\log \mathrm{age}_{\rm model} - \log \mathrm{age}_{\rm star}| < 0.2, \nonumber\\
&|(G_{\rm BP}-G_{\rm RP})_{\rm model} - (G_{\rm BP}-G_{\rm RP})_{\rm star}| < 0.1, \nonumber\\
&|M_{G,{\rm model}} - M_{G,{\rm star}}| < 0.2
\end{align}

We retrieved the models satisfying above criteria and computed their median luminosities, which was taken as the bolometric luminosity of the star. The resulting X-ray activity and stellar parameters are listed in Table~\ref{xray_results.tab}.

\begin{table*}[htp]
\centering
\caption{The basic parameters and results of the sample.}
\label{xray_results.tab}
\resizebox{\textwidth}{!}{
\begin{tabular}{rrrrrrllcclcc}
\hline\hline
\textit{Gaia} DR3 ID & RA & DEC & log Age & Dist & $A_{\rm V}$ & Class & Inst. & SpT & $N_{\rm obs}$ & $\log L_{\rm X}$ & $\log R_{\rm X}$ & Flag\\
 & ($\deg$) & ($\deg$) &  & (pc) & (mag) & & & & & & &  \\
\midrule
1003386484417913344 & 102.71681 & 60.72123 & 9.06 & 299 & 0.22 & Field & XMM & F3 & 1 & $29.08\pm 0.14$ & $-5.13\pm 0.14$ & 1 \\
1003387618288618368 & 102.60926 & 60.77658 & 9.31 & 340 & 0.25 & Field & XMM & F5 & 1 & $28.92\pm 0.15$ & $-5.06\pm 0.15$ & 1 \\
1019346784750970624 & 139.31064 & 51.61746 & 9.80 & 412 & 0.05 & Field & XMM & G1 & 1 & $29.51\pm 0.07$ & $-4.26\pm 0.07$ & 1 \\
1021434418030222208 & 144.73019 & 54.52166 & 9.55 & 137 & 0.00 & Field & XMM & K9 & 1 & $28.25\pm 0.12$ & $-4.34\pm 0.12$ & 1 \\
1023668556938986368 & 137.42459 & 54.32800 & 9.21 & 338 & 0.05 & Field & XMM & F5 & 1 & $28.73\pm 0.07$ & $-5.56\pm 0.07$ & 1 \\
1024149279742104704 & 139.90024 & 55.26462 & 9.00 & 188 & 0.03 & Field & XMM & K2 & 1 & $28.43\pm 0.06$ & $-4.62\pm 0.06$ & 1 \\
1024156980619692800 & 139.96637 & 55.40884 & 9.58 & 321 & 0.09 & Field & XMM & G9 & 1 & $28.25\pm 0.15$ & $-5.09\pm 0.15$ & 1 \\
1024711031400698112 & 143.54593 & 55.19535 & 9.10 & 368 & 0.06 & Field & XMM & F5 & 2 & $28.83\pm 0.09$ & $-5.35\pm 0.10$ & 1 \\
1027677101455282432 & 129.37445 & 50.87075 & 9.19 & 342 & 0.19 & Field & XMM & K1 & 1 & $28.47\pm 0.15$ & $-4.71\pm 0.15$ & 1 \\
1031824046998352896 & 123.61567 & 53.43217 & 9.67 & 103 & 0.00 & Field & XMM & K6 & 1 & $27.74\pm 0.10$ & $-4.89\pm 0.10$ & 1 \\
1046040491827670528 & 148.66936 & 56.49356 & 9.29 & 255 & 0.03 & Field & XMM & F9 & 1 & $29.02\pm 0.07$ & $-4.76\pm 0.07$ & 1 \\
1046051315145264512 & 148.38171 & 56.57969 & 9.70 & 206 & 0.03 & Field & XMM & K1 & 1 & $28.39\pm 0.14$ & $-4.99\pm 0.14$ & 1 \\
1083081354941073920 & 117.43737 & 59.75550 & 9.34 & 902 & 0.09 & Field & XMM & K4 & 1 & $29.45\pm 0.14$ & $-3.46\pm 0.14$ & 1 \\
1150308553880004992 & 113.87111 & 85.72440 & 9.87 & 1166 & 0.31 & Field & XMM & G7 & 1 & $29.63\pm 0.13$ & $-3.70\pm 0.13$ & 1 \\
1153359969228336256 & 225.20740 & 1.57401 & 9.08 & 412 & 0.22 & Field & XMM & F3 & 2 & $28.80\pm 0.14$ & $-5.39\pm 0.13$ & 1 \\
1153395875154938112 & 225.46778 & 1.88178 & 9.15 & 446 & 0.19 & Field & XMM & F5 & 1 & $29.50\pm 0.15$ & $-4.78\pm 0.15$ & 1 \\
1153404125787788416 & 225.20847 & 1.88751 & 9.58 & 113 & 0.00 & Field & XMM & M2 & 2 & $27.89\pm 0.09$ & $-4.24\pm 0.07$ & 1 \\
1153526309017518208 & 226.73986 & 1.65767 & 9.74 & 189 & 0.34 & Field & XMM & G5 & 5 & $28.58\pm 0.06$ & $-5.04\pm 0.06$ & 1 \\
1154089602568213504 & 223.93888 & 1.68732 & 9.11 & 318 & 0.19 & Field & XMM & F5 & 1 & $29.44\pm 0.09$ & $-4.77\pm 0.09$ & 1 \\
1154985154788975104 & 223.72720 & 3.59286 & 9.29 & 184 & 0.09 & Field & XMM & K3 & 1 & $28.16\pm 0.10$ & $-4.75\pm 0.10$ & 1 \\
\bottomrule
\end{tabular}}
\vspace{0.05cm}
\begin{flushleft}
\footnotesize
\textbf{Notes.} 
Class: ``OC'' denotes open cluster members, while ``Field'' denotes field stars. 
SpT: Spectral types estimated from the $G_{\rm BP}-G_{\rm RP}$ color following \citet{2013ApJS..208....9P}. 
$N_{\rm obs}$: The observations number for each source.
Flag: 1 represents X-ray detections, while 0 represent X-ray upper limits. 
\end{flushleft}

\end{table*}

\subsection{Upper Limits}
\label{upperlimit.sec}

To account for X-ray non-detections, we constructed a supplementary sample of stars that were covered by \textit{XMM-Newton} observations but were not detected in the X-ray source catalog. 
We estimated their upper limits using the RapidXMM upper-limit database \citep{2022MNRAS.511.4265R}, which provides pre-computed upper limits based on aperture photometry of \textit{XMM-Newton} observations.

Rather than deriving upper limits for the whole OC and field sample, which includes millions of stars, we performed a preliminary selection to save computational time.
We identified all \textit{XMM-Newton} observations associated with the detected sources in our sample and obtained the pointing center coordinates of each observation using \texttt{astroquery}. For each observation ID, we cross-matched the pointing center with the OC and field star catalogs within a radius of $10\arcmin$. Stars falling within this radius were considered covered by the corresponding \textit{XMM-Newton} observation. Upper limits for these sources were then obtained using RapidXMM.

To ensure reliable upper-limit measurements, we retained only those with ``pointed" observation type (rather than the ``slew" type), valid quality flags (${\tt band8\_flags}=0$), and exposure times longer than 10 ks. 
Furthermore, we make a cleaning process to exclude some extremely high and abnormal upper limits (see Appendix \ref{upperlimit_select.sec}).

The EPIC-pn band-8 ($0.2$--$12$ keV) $1\sigma$ count-rate upper limits, returned by RapidXMM, were converted to unabsorbed $0.1$--$2.4$ keV fluxes using \texttt{PIMMS}, adopting the same APEC model as used for the detected sources. The corresponding $L_{\rm X}$ and $R_{\rm X}$ upper limits were then calculated following the same procedure as for the detected sample.
Finally, we used the same selection criteria to select dwarfs with upper-limit estimates as detected stars in Section~\ref{sample_selection.sec}.

\begin{figure*}[!htbp]
\centering
    \includegraphics[width=0.8\textwidth]{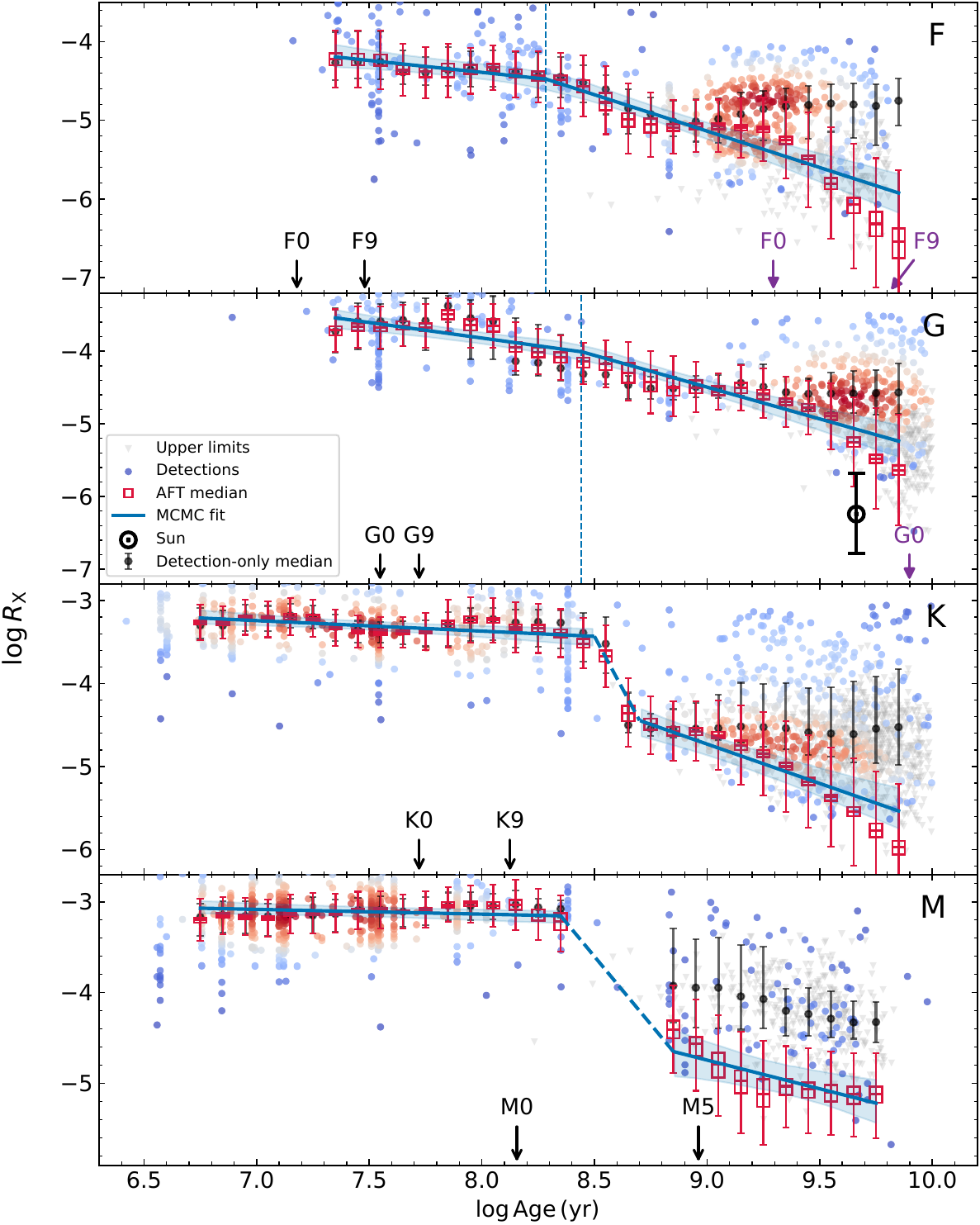}
    \caption{The relation between $R_{\rm X}$ and stellar age.  
    The colored points represent the number density of detected sources, while gray triangles indicate upper limits.
    The black points are the median values of  $\log R_{\rm X}$ in each age bin, with the 16th-84th percentiles shown as vertical error bars, calculated from detected sources only.
    Within each red box, the horizontal line marks the median $\log R_{\rm X}$ inferred from the LogNormal AFT analysis in that age bin, with the 16th-84th percentiles shown as vertical error bars.
    The red open boxes represent the 16th-84th percentile range of the median  $\log R_{\rm X}$ values derived from 100 bootstrap resamplings, serving to indicate the stability of the LogNormal AFT analysis.
    The blue lines represent the linear fits derived from MCMC analysis of the red lines in boxes, with the blue shaded regions indicating the $1\sigma$ uncertainties.
    Dashed blue lines mark regions where no fit was applied. The black and purple arrows indicate, respectively, the times when stars enter and leave the main sequence. The Sun is marked with an open circle. For clarity, only the 5th-95th percentile range of the data is shown in the figure, although all sources were included in the analysis.}
    \label{distribution_xray_log_fit_mcmc.fig}
\end{figure*}

\begin{figure*}[!htbp]
\centering
    \includegraphics[width=0.9\textwidth]{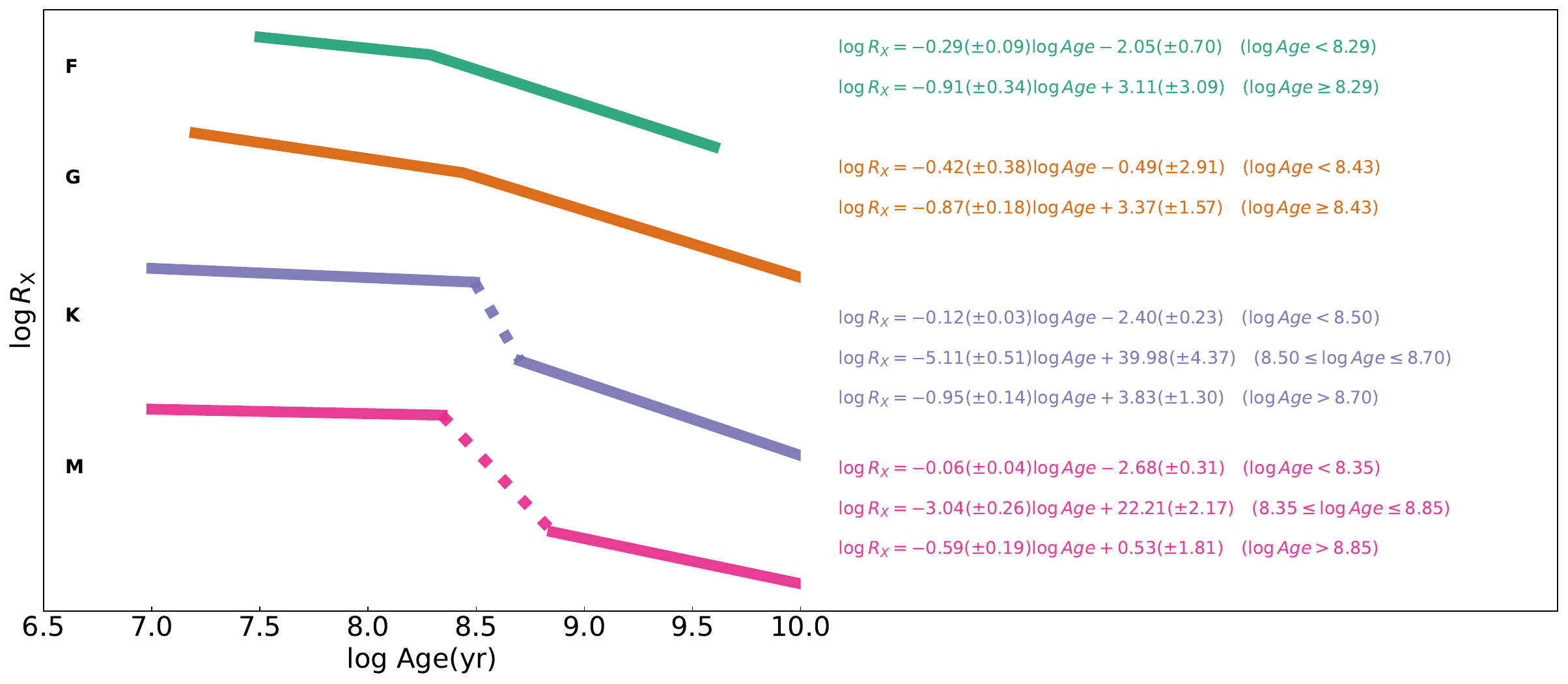}
    \caption{Schematic view of the relation between $R_{\rm X}$ and stellar age based on the MCMC results.}
    \label{xray_evo_model.fig}
\end{figure*}

\begin{table*}
\centering
\caption{Piecewise linear fit parameters for the age evolution of $\log L_{\rm X}$, $\log R_{\rm X}$, and $\log L_{\rm XUV}$. The relations are expressed as $y = a t + b$, where $t=\log{\rm Age}$ with age in yr.}
\label{fits_results.tab}
\renewcommand{\arraystretch}{1.2}
\begin{tabular}{llcccccccc}
\hline\hline
Observable & SpT 
& $a_1$ & $b_1$ & $t_1$ 
& $a_2$ & $b_2$ & $t_2$ 
& $a_3$ & $b_3$ \\
\hline
\multirow{4}{*}{$\log R_{\rm X}$}
 & F & -- & -- & -- & $-0.29\pm 0.19$ & $-2.05\pm 1.54$ & 8.29 & $-0.91\pm 0.21$ & $3.11\pm 2.85$ \\
 & G & -- & -- & -- & $-0.42\pm 0.17$ & $-0.49\pm 1.37$ & 8.43 & $-0.87\pm 0.19$ & $3.37\pm 2.54$ \\
 & K & $-0.12\pm 0.09$ & $-2.40\pm 0.66$ & 8.50 & $-5.11$ & $39.98$ & 8.70 & $-0.95\pm 0.27$ & $3.83\pm 2.45$ \\
 & M & $-0.06\pm 0.06$ & $-2.68\pm 0.42$ & 8.35 & $-3.04$ & $22.21$ & 8.85 & $-0.59\pm 0.44$ & $0.53\pm 4.08$ \\

\hline
\multirow{4}{*}{$\log L_{\rm X}$}
 & F & -- & -- & -- & $-0.22\pm 0.17$ & $31.31\pm 1.35$ & 8.32 & $-0.90\pm 0.21$ & $37.03\pm 2.63$ \\
 & G & -- & -- & -- & $-0.37\pm 0.19$ & $32.65\pm 1.47$ & 8.43 & $-1.01\pm 0.21$ & $38.10\pm 2.79$ \\
 & K & $-0.45\pm 0.12$ & $33.31\pm 0.90$ & 8.50 & $-4.83$ & $70.53$ & 8.70 & $-1.05\pm 0.28$ & $37.65\pm 2.57$ \\
 & M & $-0.31\pm 0.14$ & $31.63\pm 1.05$ & 8.35 & $-2.21$ & $47.49$ & 8.85 & $-0.99\pm 0.27$ & $36.62\pm 2.52$ \\

\hline
\multirow{4}{*}{$\log L_{\rm XUV}$}
 & F & -- & -- & -- & $-0.19\pm 0.15$ & $31.41\pm 1.23$ & 8.31 & $-0.69\pm 0.16$ & $35.55\pm 2.24$ \\
 & G & -- & -- & -- & $-0.38\pm 0.17$ & $32.97\pm 1.36$ & 8.44 & $-0.79\pm 0.17$ & $36.47\pm 2.47$ \\
 & K & $-0.47\pm 0.10$ & $33.56\pm 0.75$ & 8.50 & $-3.50$ & $59.32$ & 8.70 & $-0.93\pm 0.29$ & $37.01\pm 2.70$ \\
 & M & $-0.40\pm 0.13$ & $32.44\pm 0.98$ & 8.35 & $-2.60$ & $50.83$ & 8.85 & $-0.25\pm 0.47$ & $30.06\pm 4.37$ \\
\hline
\end{tabular}

\vspace{0.05cm}
\begin{flushleft}
\footnotesize
\textbf{Notes.}
The parameters $(a_i, b_i)$ describe the slope and intercept of each linear segment, 
while $t_i$ denotes the logarithmic break time separating adjacent age regimes. 
For K- and M-type stars, the intermediate segment connects the two fitted outer segments continuously.
\end{flushleft}
\end{table*}

\begin{figure}[t]
    \centering
    \includegraphics[width=0.48\textwidth]{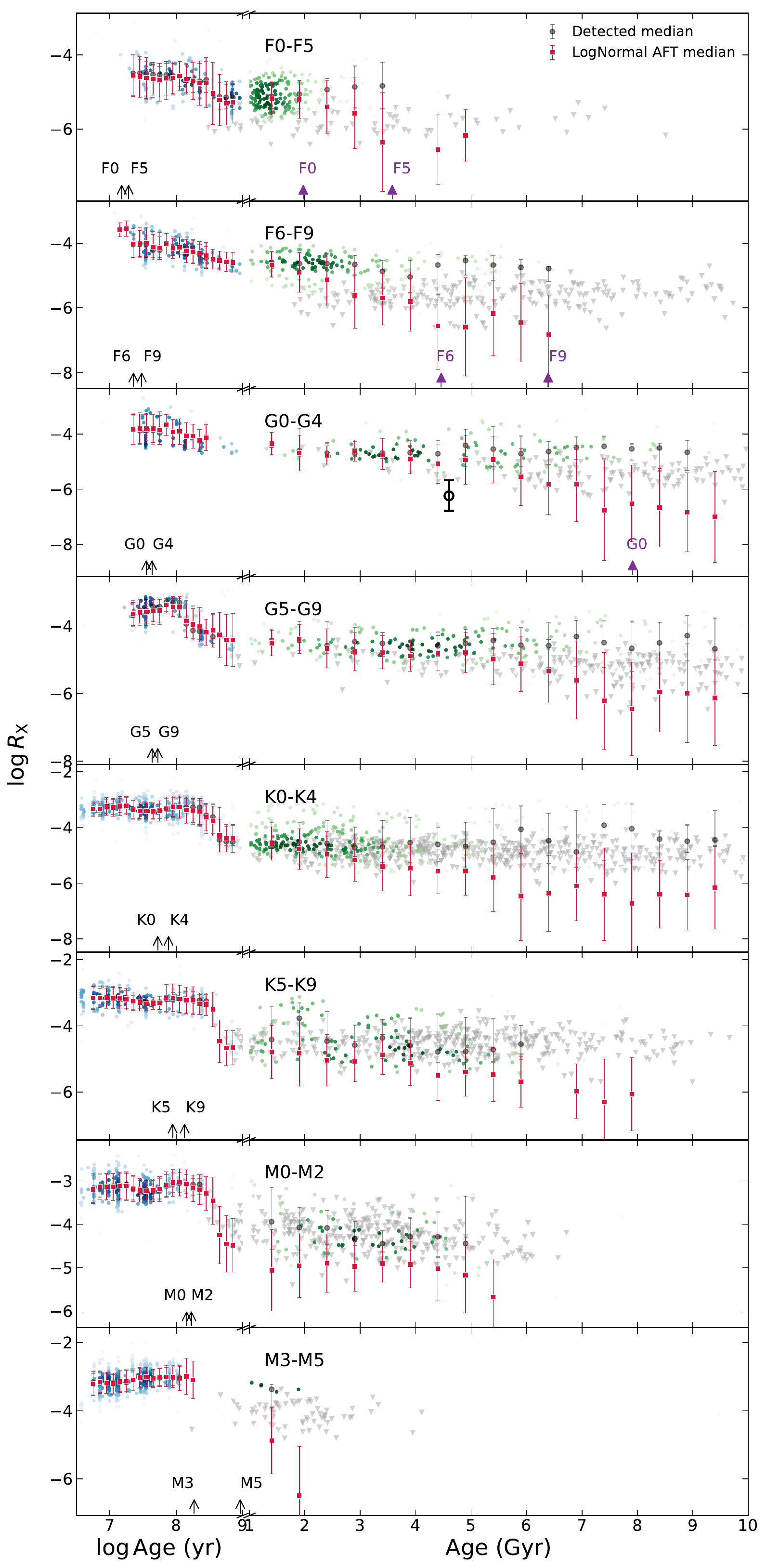}
    \caption{Distribution of $\log R_{\rm X}$ as a function of stellar age for different spectral-type bins. Blue and green density maps show X-ray detections at ages younger and older than 1 Gyr, respectively. Gray triangles are upper limits.
    Black circles and error bars show the median and 16th--84th percentile range of the detected sources. Red squares and error bars show the median and 16th--84th percentile range estimated with a LogNormal AFT model using both detections and upper limits. The arrows indicate the approximate main-sequence lifetime boundaries of the corresponding spectral subtypes.} \label{distribution_rx_all.fig}
\end{figure}

Before investigating the activity evolution, we applied a distance cutoff to stars with upper-limit estimates, owing to their huge number. For each spectral type, we calculated the peak distance of the kernel density estimation (KDE) using the detected OC sample as the completeness limit, and applied this limit to the upper-limit stars. The adopted distance limits are 429 pc for F-type stars, 434 pc for G-type stars, 429 pc for K-type stars, and 390 pc for M-type stars (Figure~\ref{oc_dist_distribution.fig}). 
This cutoff was applied only to upper-limit stars; detected stars were not subject to this restriction, as their number is much smaller than that of the upper-limit sources.

\section{The x-ray activity evolution}
\label{evo.sec}

We used the $R_{\rm X}$ ($\approx L_{\rm X}/L_{bol}$) to represent the X-ray activity. 
For each spectral type, we computed median $R_{\rm X}$ values in stellar-age bins. 
Starting from a logarithmic age of 6.5, we constructed overlapping age bins with a width of 0.5~dex. The bin center was shifted by 0.1~dex at each step to produce a moving-median sequence. Only bins containing at least 20 stars were retained. 

Since the data contain limits, a survival function should be used to account for the limits \citep{1985ApJ...293..192F}. 
However, in several old-age field-star bins, the number of upper limits substantially exceeds that of detections, resulting in heavily censored samples (Figure~\ref{detected_to_all.fig}). In such cases, the Kaplan--Meier estimator, as previously used in several studies \citep{2016ApJ...830...44N,2023ApJ...951...44R,2025A&A...694A..93Z} may not robustly constrain the median value. We therefore adopted a log-normal accelerated failure time (AFT) model, a parametric survival-analysis method, to estimate the median log$R_{\rm X}$ by fitting the censored distribution using \textit{lifelines} package \citep{Davidson-Pilon2019}.

The X-ray activity evolution is shown in Figure~\ref{distribution_xray_log_fit_mcmc.fig}.
The median values after considering the upper limits and uncertainties were then used for a piecewise curve fitting. The best-fit parameters from this step served as initial values for a subsequent MCMC fit, which provided the final age evolution trends for each spectral type. During the fitting process, for F- and G-type stars, due to the lack of young stars, we performed only a two-phase fit. For K- and M-type stars, because of the lack of stars during log age $\sim$8.5–8.7, we did not perform a continuous fit. Instead, we performed separate linear fits for the young and old populations. The Sun was also plotted with an X-ray luminosity range of $L_{\rm X}=10^{26.8}$–$10^{27.9}$ erg s$^{-1}$ from \citet{2003ApJ...593..534J} and $\log R_{\rm X} = -6.78- -5.68$.

To assess the robustness of the median activity--age relation inferred from the censored data, we performed a bootstrap stability test for the LogNormal AFT analysis. For all spectral types, in each age bin, we resampled the sources using the original sample (including detections and upper limits) with replacement 100 times and refit the LogNormal AFT model to estimate the median $\log R_{\rm X}$. The resulting bootstrap distributions provide an empirical estimate of the uncertainty arising from sample selection and censoring. The 100 median values are shown as the red bars in Figure~\ref{distribution_xray_log_fit_mcmc.fig}, where the scatter is very small, supporting the robustness of the inferred activity evolution. 

In addition, we assessed the impact of age uncertainties on the inferred activity--age relations, as we did not include age uncertainties in our fitting procedure. We performed a Monte Carlo test in which we resampled the stellar ages by adding Gaussian noise according to the uncertainties reported in \citet{2025ApJS..280...13W}, and then refitted the activity--age relation using the resampled age set. This process was repeated 100 times. The resulting relations shown in Figure~\ref{age_error_fit.fig} are nearly indistinguishable from the original relations in Figure \ref{distribution_xray_log_fit_mcmc.fig}, suggesting that age uncertainties have negligible effects on the evolutionary trends.

For each spectral type, $R_{\rm X}$ decreases by approximately 2--3 orders of magnitude from young to old ages. 
For F- and G-type stars, because the sample lacks pre–MS stars, our results only trace their activity evolution after they enter the MS. In the log-log plane, the observed trend exhibits a relatively flat/shallow slope during the early MS phase and becomes steeper at later ages. 
For K and M stars, our sample lacks observational data during log age $\sim$8.5--8.7. Therefore, we only observe the saturated and slow-decline segment in our data. However, based on the early saturated phase and the late slow-decline phase, we inferred that this is a very rapid decline phase between them. 
An F-test comparison provides statistical support for the three-phase description over the two-phase model, with $p=1.96\times10^{-4}$ for K-type stars and $p=1.97\times10^{-7}$ for M-type stars. 

To clearly compare the evolutionary trends across different spectral types, we plotted the $R_{\rm X}$-age relations derived from our MCMC fits (as shown in Figure~\ref{xray_evo_model.fig}). 
The best-fit $\log L_{\rm X}$ and $\log R_{\rm X}$ evolutionary parameters for different spectral types are listed in Table~\ref{fits_results.tab}. 
We also plotted the field-star sample on a linear age scale in Figure~\ref{distribution_rx_all.fig} to better visualize the activity–age relation at old ages.

\subsection{Comparison with Previous Studies}
\label{other_work.sec}

\begin{figure*}[!htbp]
\centering
    \includegraphics[width=0.9\textwidth]{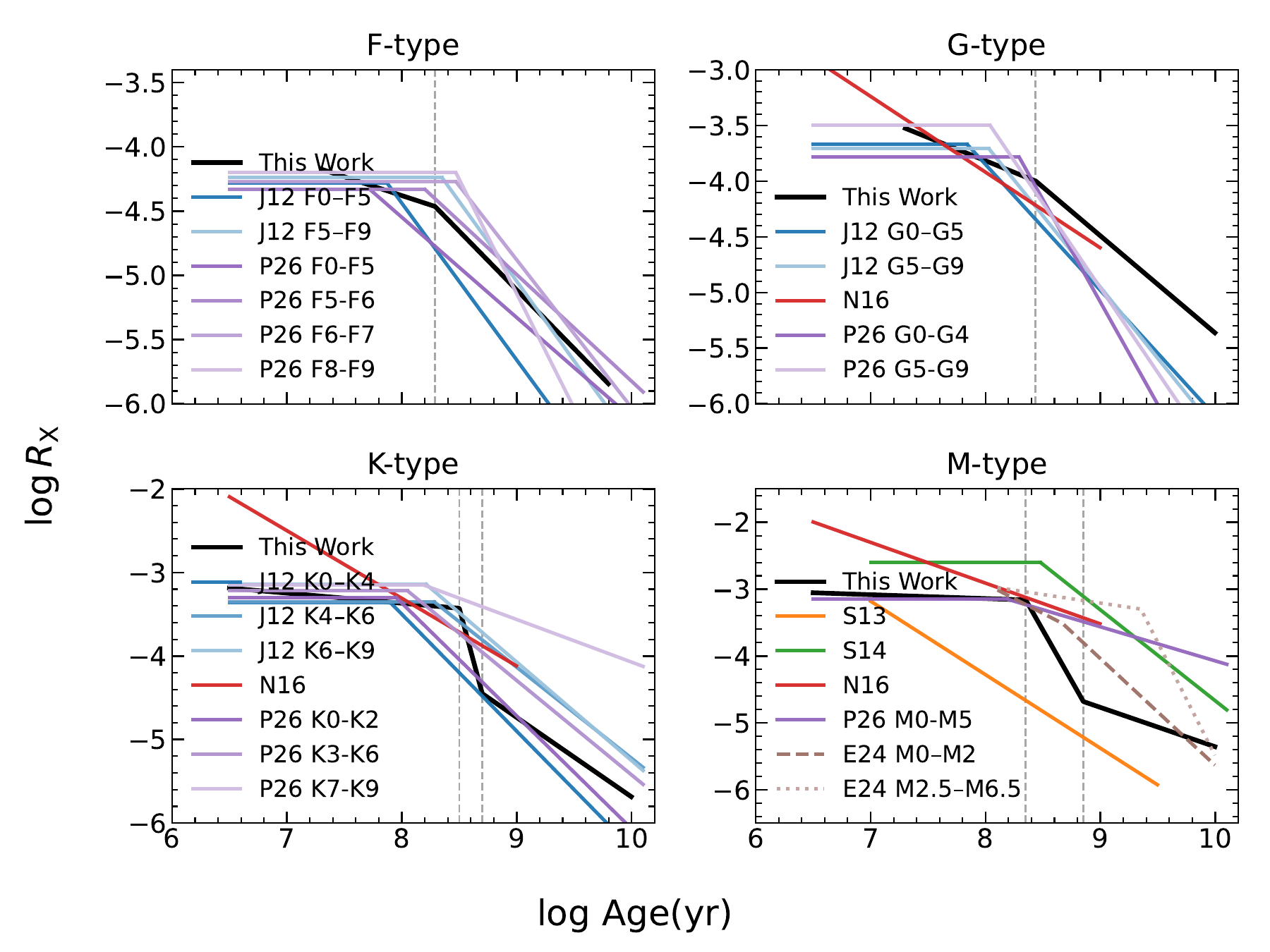}
    \caption{Comparison of the $\log R_{\rm X}$ evolutionary relations derived in this work with those in previous studies. The solid black line represents the best-fit relation obtained in this work. Colored lines show literature results from \citet{2012MNRAS.422.2024J} (J12), \citet{2013MNRAS.431.2063S} (S13), \citet{2014AJ....148...64S} (S14), \citet{2016ApJ...830...44N} (N16), \citet{2024ApJ...960...62E} (E24), and \citet{2026Pezzotti} (P26).}
    \label{rx_relation_other_paper.fig}
\end{figure*}

We compiled previous studies on the $R_{\rm X}$--age relations, including \citet{2012MNRAS.422.2024J}, \citet{2013MNRAS.431.2063S},\citet{2014AJ....148...64S}, \citet{2016ApJ...830...44N}, \citet{2024ApJ...960...62E}, and \citet{2026Pezzotti}. 
For comparison, we summarized these literature relations in Figure~\ref{rx_relation_other_paper.fig} and list their corresponding slopes in Table~\ref{slope_age.tab}.

Overall, our results are broadly consistent with earlier studies. For F- and G-type stars, our relations lie close to those reported by \citet{2012MNRAS.422.2024J} and \citet{2026Pezzotti}. One small difference is that we did not observe a clear saturated phase for F stars, since our sample lacks pre-MS F stars.

The most notable difference is seen for K- and M-type stars, whose activity evolution follows a three-phase relation, in contrast to the single-slope decline 、or two-phase prescriptions adopted in previous studies \citep{2012MNRAS.422.2024J, 2013MNRAS.431.2063S, 2014AJ....148...64S, 2016ApJ...830...44N, 2024ApJ...960...62E, 2026Pezzotti}. This discrepancy likely arises because previous studies relied on only a small number of old stars with age measurements to constrain the relation.
It is noteworthy that \citet{2024ApJ...960...62E} included a population of stars older than  1 Gyr and performed a two-phase model fit (see their Figure~3). However, the early-M stars in their sample exhibit large scatter, particularly at old ages. Our evolutionary curve can provide a good fit to their data as well.

\begin{figure*}[!htbp]
\centering
    \includegraphics[width=0.9\textwidth]{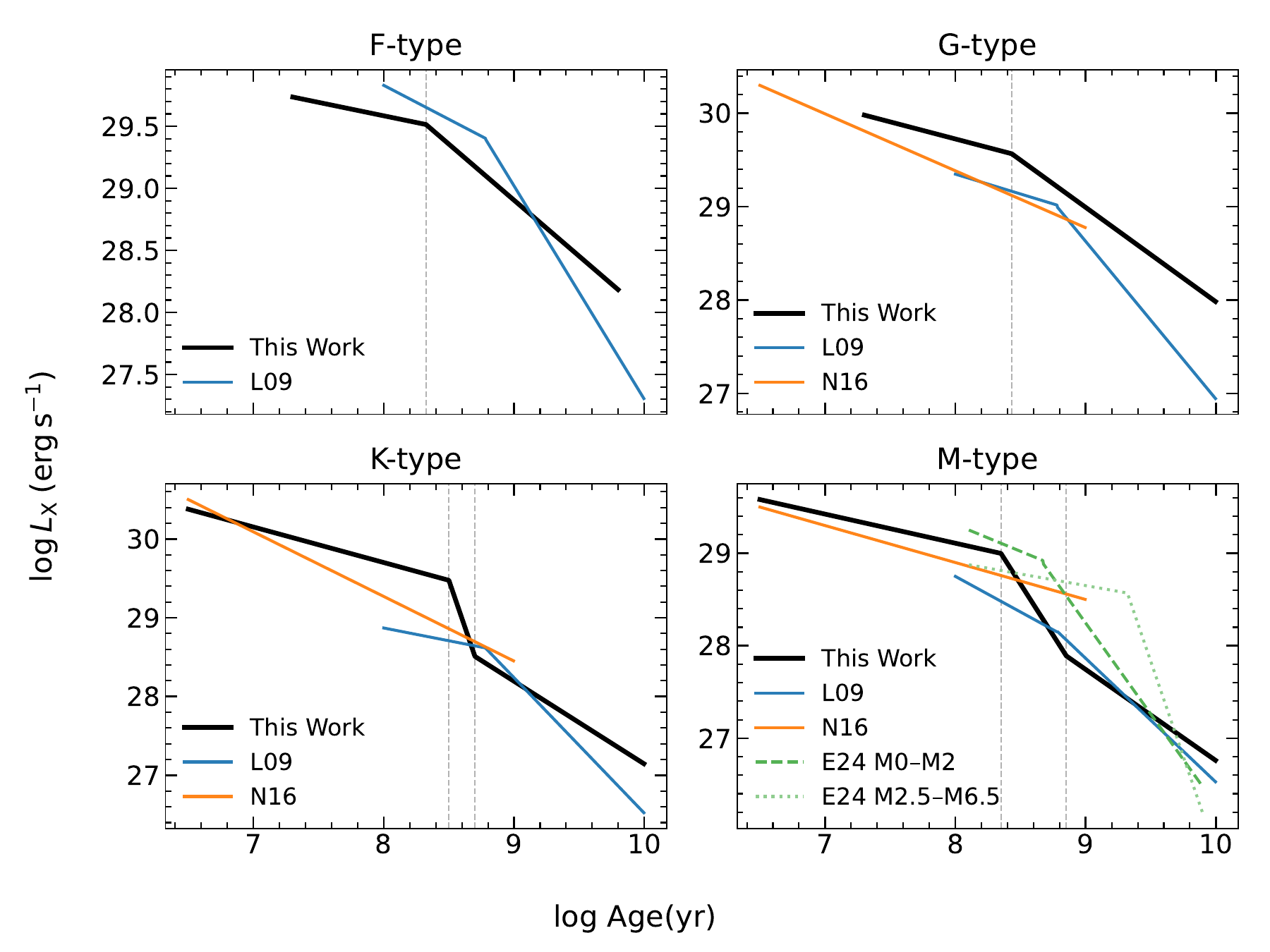}
    \caption{Comparison of the $\log L_{\rm X}$ evolutionary relations derived in this work with those in previous studies. The solid black line represents the best-fit relation obtained in this work. Colored lines show literature results from \citet{2009A&A...506..399L} (L09), \citet{2016ApJ...830...44N} (N16), and \citet{2024ApJ...960...62E} (E24).}
    \label{lx_relation_other_paper.fig}
\end{figure*}

We also presented a compilation of the $L_{\rm X}$--age relations reported in the literature, including \citet{2009A&A...506..399L}, \citet{2016ApJ...830...44N}, and \citet{2024ApJ...960...62E}.
Our relations are broadly consistent with those studies, as shown in Figure~\ref{lx_relation_other_paper.fig}.

\begin{table*}
\caption{Comparison of break ages and slopes in the $\log R_{\rm X}$--Age relation for different spectral types in the different studies.}
\label{slope_age.tab}
\noindent

\resizebox{\textwidth}{!}{
\begin{tabular}{llcccccccc}
\hline\hline
Index & Study & F-type & log Age(yr) & G-type & log Age(yr) & K-type & log Age(yr) & M-type & log Age(yr) \\
\hline
\multirow{13}{*}{$\log R_{\rm X}$}
  & \citet{2012MNRAS.422.2024J} & $[-1.24, -1.22]$ & $\geq 7.87$ & $[-1.28, -1.13]$ & $\geq 7.84$ & $[-1.4, -1.09]$ & $\geq 7.90$ & -- & -- \\
  & \citet{2014AJ....148...64S} & -- & -- & -- & -- & -- & -- & $-1.36\pm 0.32$ & $\geq 8.48$ \\
  & \citet{2016ApJ...830...44N} & -- & -- & $-0.68\pm 0.12$ & $[6.5, 9]$ & $-0.81\pm 0.19$ & $[6.5, 9]$ & $-0.61\pm 0.12$ & $[6.5, 9]$ \\
  \cline{2-10}
    & \multirow{2}{*}{\citet{2024ApJ...960...62E}} 
  & -- & -- 
  & -- & -- 
  & -- & -- 
  & $-0.88\pm0.24$ (M0--M2) / $-0.25\pm0.16$ (M2.5--M6.5) 
  & $\leq 8.48 / 9.36$ \\

  & 
  & -- & -- 
  & -- & -- 
  & -- & -- 
  & $-1.60\pm0.26$ (M0--M2) / $-3.40\pm0.20$ (M2.5--M6.5) 
  & $\geq 8.48 / 9.36$ \\
 \cline{2-10}
   & \multirow{5}{*}{\citet{2026Pezzotti}} 
  & $-0.78\pm 0.12$ (F0--F5) & $\geq 7.72$ 
  & $-1.84\pm 0.23$ (G0--G4) & $\geq 8.29$ 
  & $-1.35\pm 0.15$ (K0--K2) & $\geq 7.95$ 
  & $-0.51\pm 0.26$ (M0--M5) & $\geq 8.19$ \\

  & 
  & $-0.83\pm 0.26$ (F5--F6) & $\geq 8.20$ 
  & $-1.52\pm 0.30$ (G5--G9) & $\geq 8.04$ 
  & $-1.13\pm 0.20$ (K3--K6) & $\geq 8.05$ 
  & -- & -- \\

  & 
  & $-1.15\pm 0.22$ (F6--F7) & $\geq 8.47$ 
  & -- & -- 
  & $-0.51\pm 0.26$ (K7--K9) & $\geq 8.19$ 
  & -- & -- \\

  & 
  & $-1.78\pm 0.29$ (F8--F9) & $\geq 8.47$ 
  & -- & -- 
  & -- & -- 
  & -- & -- \\
  &
  & $-1.65\pm 0.11$ (All) & $\geq 8.15$ 
  & $-1.65\pm 0.11$ (All) & $\geq 8.15$ 
  & $-1.65\pm 0.11$ (All) & $\geq 8.15$ 
  & $-1.65\pm 0.11$ (All) & $\geq 8.15$ \\
 \cline{2-10}
  & \multirow{3}{*}{This Work} 
  & $-0.29\pm 0.19$ & $\le 8.29$ 
  & $-0.42\pm 0.17$ & $\le 8.43$ 
  & $-0.12\pm 0.09$ & $\le 8.50$ 
  & $-0.06\pm 0.06$ & $\le 8.35$ \\
  & 
  & -- & -- 
  & -- & -- 
  & $-5.11$ & $[8.50, 8.70]$ 
  & $-3.04$ & $[8.35, 8.85]$ \\
  & 
  & $-0.91\pm 0.21$ & $> 8.29$ 
  & $-0.87\pm 0.19$ & $> 8.43$ 
  & $-0.95\pm 0.27$ & $> 8.70$ 
  & $-0.59\pm 0.44$ & $> 8.85$ \\
 \hline

\multirow{9}{*}{$\log L_{\rm X}$}
  & \multirow{2}{*}{\citet{2009A&A...506..399L}} & $-0.547$ & $\le 8.78$ & 0.425 & $\le 8.78$ & $-0.324$ & $\le 8.78$ & $-0.77$ & $\le 8.78$ \\
   & & $-1.72$ & $> 8.78$ & $-1.69$ & $> 8.78$ & $-1.72$ & $> 8.78$ & $-1.34$ & $> 8.78$ \\
  \cline{2-10}
  & \citet{2013MNRAS.431.2063S} & -- & -- & -- & -- & -- & -- & $-1.10\pm 0.02$ & $\geq 7$ \\
  & \citet{2016ApJ...830...44N} & -- & -- & $-0.61\pm 0.12$ & $[6.5, 9]$ & $-0.82\pm 0.16$ & $[6.5, 9]$ & $-0.40\pm 0.17$ & $[6.5, 9]$ \\
  \cline{2-10}
    & \multirow{2}{*}{\citet{2024ApJ...960...62E}} 
  & -- & -- 
  & -- & -- 
  & -- & -- 
  & $-0.58\pm0.24$ (M0--M2) / $-0.25\pm0.16$ (M2.5--M6.5) 
  & $\leq 8.48 / 9.36$ \\

  & 
  & -- & -- 
  & -- & -- 
  & -- & -- 
  & $-0.72\pm0.27$ (M0--M2) / $-3.14\pm0.20$ (M2.5--M6.5) 
  & $\geq 8.48 / 9.36$ \\
 \cline{2-10}
  & \multirow{3}{*}{This Work} 
  & $-0.22\pm 0.17$ & $\le 8.32$ 
  & $-0.37\pm 0.19$ & $\le 8.43$ 
  & $-0.45\pm 0.12$ & $\le 8.50$ 
  & $-0.31\pm 0.14$ & $\le 8.35$ \\
  & 
  & -- & -- 
  & -- & -- 
  & $-4.83$ & $[8.50, 8.70]$ 
  & $-2.21$ & $[8.35, 8.85]$ \\
  & 
  & $-0.90\pm 0.21$ & $> 8.32$ 
  & $-1.01\pm 0.21$ & $> 8.43$ 
  & $-1.05\pm 0.28$ & $> 8.70$ 
  & $-0.99\pm 0.27$ & $> 8.85$ \\
  \hline

\multirow{2}{*}{$\log \frac{L_{\rm X}}{(R/R_{\odot})^2}$}
  & \citet{2017MNRAS.471.1012B} & $-2.80\pm 0.72$ & $[9, 10]$ & $-2.80\pm 0.72$ & $[9, 10]$ & $-2.80\pm 0.72$ & $[9, 10]$ & -- & -- \\
  & \citet{2025PASP..137k4206A} & $[-2.11, -1.37]$ & $[9, 10]$ & $[-2.11, -1.37]$ & $[9, 10]$ & $[-2.11, -1.37]$ & $[9, 10]$ & -- & -- \\
\hline
\end{tabular}
}
\begin{flushleft}
\footnotesize
\textbf{Notes.} 
$\log \frac{L_{\rm X}}{(R/R_{\odot})^2}$ represents the X-ray luminosity normalized by the stellar surface area, serving as a proxy for the surface X-ray flux.
\end{flushleft}
\end{table*}

\subsection{Old Stars with High X-ray Activity}
\label{old_stars.sec}

We found some field stars exhibiting unusually high X-ray activity levels ($\log R_{\rm X}> -4$) that deviate significantly from the best-fitting evolutionary relations, particularly among K- and M-type stars. 
As shown in section~\ref{sample_selection.sec}, we have cross-matched with SIMBAD and some photometric catalogs to exclude possible contaminants (i.e., binaries, galaxies, AGNs).
To further check the possibility of unresolved multiplicity, we cross-matched these stars with large spectroscopic surveys including LAMOST, SDSS, APOGEE, RAVE, SEGUE, and DESI and examined their radial velocity (RV) variability.
Sources with RV amplitudes $\Delta{\rm RV} > 10~{\rm km~s^{-1}}$ can be identified as spectroscopic binaries. 
However, only a small number of stars have sufficient observations to probe such variability, and even fewer show large RV amplitudes.

Then, we examined whether strong X-ray flares captured during individual observations could account for the elevated X-ray luminosities of these stars. For the high-activity stars with \textit{Chandra} observations, we retrieved the individual observations and extracted light curves using the \texttt{CIAO} software package (the \textit{Chandra} Interactive Analysis of Observations). 
We generated background-subtracted light curves with a time binning of 500~s, adopting a circular source extraction region with a radius of $\sim$3.5\arcsec, and a surrounding background annulus extending from $\sim$6\arcsec\ to $\sim$11\arcsec. However, we found no clear flare event in any observation, suggesting flares are also unlikely to be the cause of the activity excess.

We further examined whether the presence of highly active old stars could be explained by age-estimation inaccuracies. \citet{2025ApJS..280...13W} showed that some stellar ages were substantially overestimated in their age catalog (see the right panel of their Figure~7; 123 out of 2236 stars, or 5.5\%, have ages more than twice their reference ages). 
In their validation sample, 2180 of the 2236 stars were assigned ages older than 1 Gyr, of which 46 (2.1\%) have reference ages below 1 Gyr.

Applying this 2.1\% misclassification rate to our 1108 field stars with ages $>$ 1 Gyr yields about 23 potentially misclassified stars.
In our sample, 178 stars are both old and highly active (age $>1$ Gyr and $\log R_{\rm X}>-4$). Even if all 23 potentially misclassified stars were highly active, they would account for only about 13\% of these 178 stars. 
Furthermore, We checked the age uncertainties of these 178 stars: 35 have uncertainties greater than 50\% and none exceed 100\%.
Thus, while age overestimation may contribute to this population, it is unlikely to explain the majority of the old, highly active stars.

\section{Discussion}
\label{disscuss.sec}

\subsection{Variability}
\label{variable.sec}
\begin{figure*}
    \centering
    \subfigure[]{
    \includegraphics[width=0.48\textwidth]{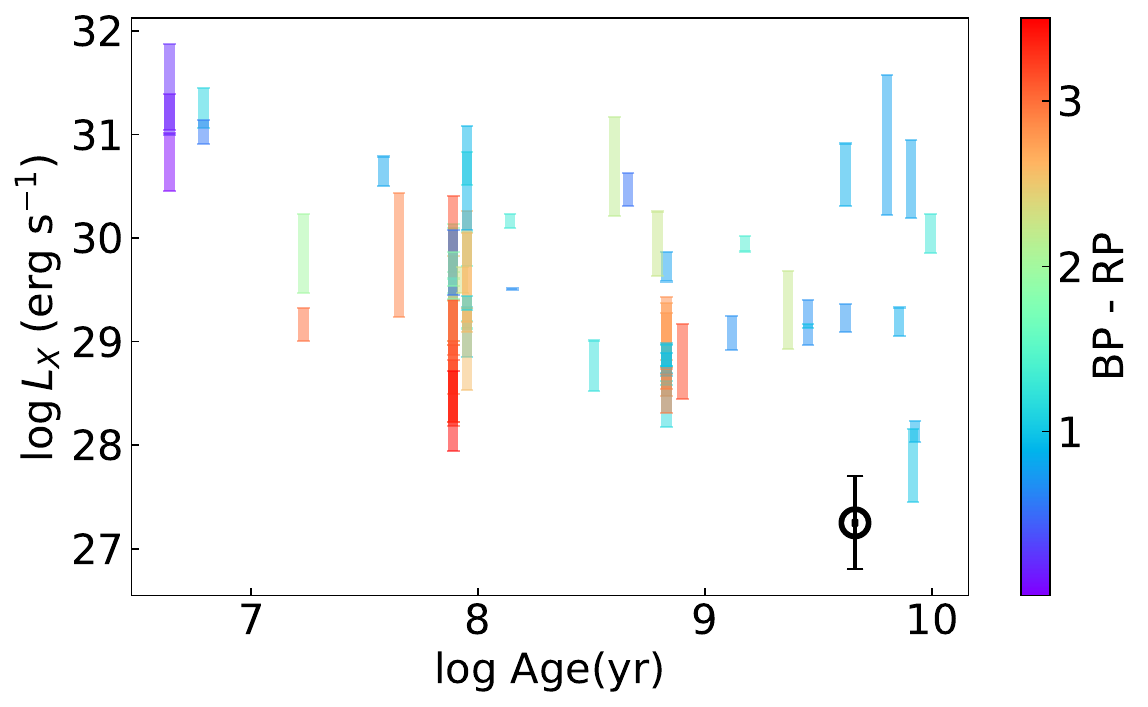}
    \label{lx_variable.fig}}
    \subfigure[]{
    \includegraphics[width=0.48\textwidth]{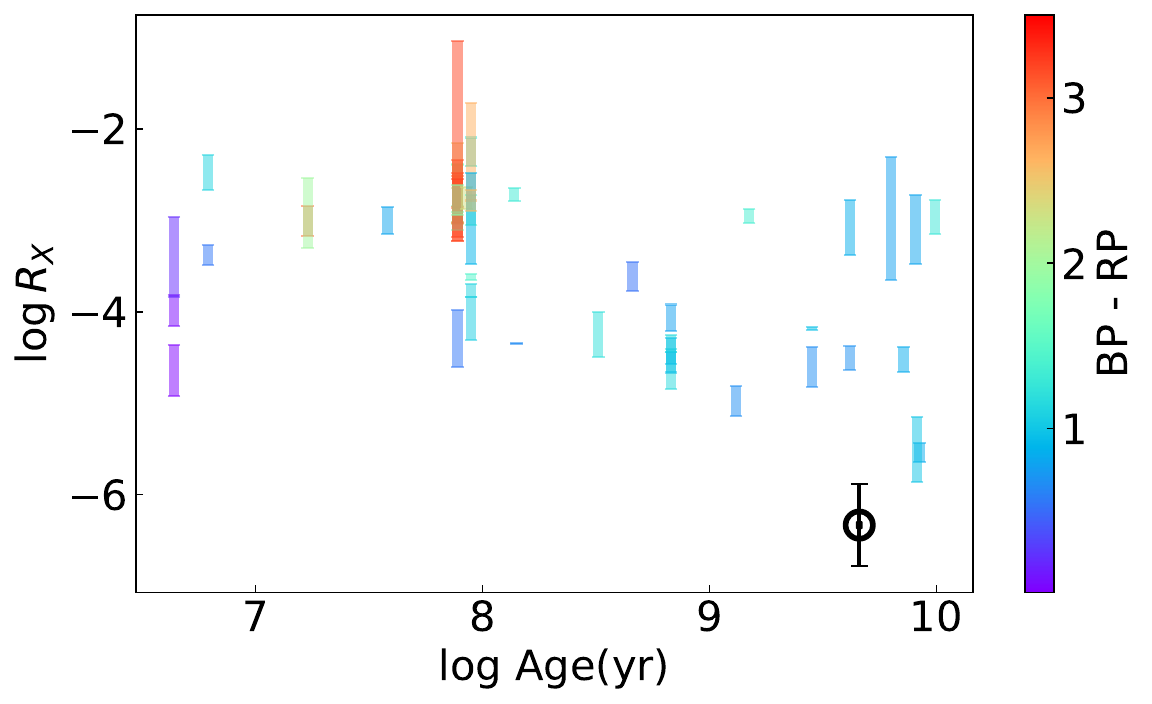}
    \label{rx_variable.fig}}
    \caption{$\log L_{\rm X}$ variability (Panel a) and $\log R_{\rm X}$ variability (Panel b) with multi-observations. The Sun is shown with an open circle. }
    \label{variable.fig}
\end{figure*}

Multiple X-ray observations introduce an additional source of variability that may influence the inferred activity--age relation. We use the \textit{Chandra} sample as a test, and found that 63 stars have multiple observations. Figure~\ref{variable.fig} shows the highest and lowest measured $\log L_{\rm X}$ (Fig. \ref{lx_variable.fig}) and $\log R_{\rm X}$ (Fig. \ref{rx_variable.fig}) of these stars. 
For comparison, the range of solar X-ray variability over its activity cycle is shown. It is interesting to note that many stars share a similar variation amplitude to the Sun, suggesting that they may also experience activity cycles.
Simultaneously, the fact that most stars show only modest variations indicates that this variability does not significantly bias our statistical trends.

There are also some stars displaying much larger variability in X-ray luminosity. Such variability is often interpreted as being associated with a flare, which can enhance coronal emission by orders of magnitude on short timescales \citep[e.g.,][]{2021NatAs...5..298C,2013ApJS..207...15K,2004A&A...416..713G}. We checked the light curves of those sources, and found seven flare events. Note that these stars with large variability are mostly young stars, with no overlap of those old high-activity outliers discussed in Section~\ref{old_stars.sec}. 

\subsection{Comparison between Sun and the Analogs}
\label{analogs.sec}

\begin{figure*}[!htbp]
\centering
    \includegraphics[width=0.98\textwidth]{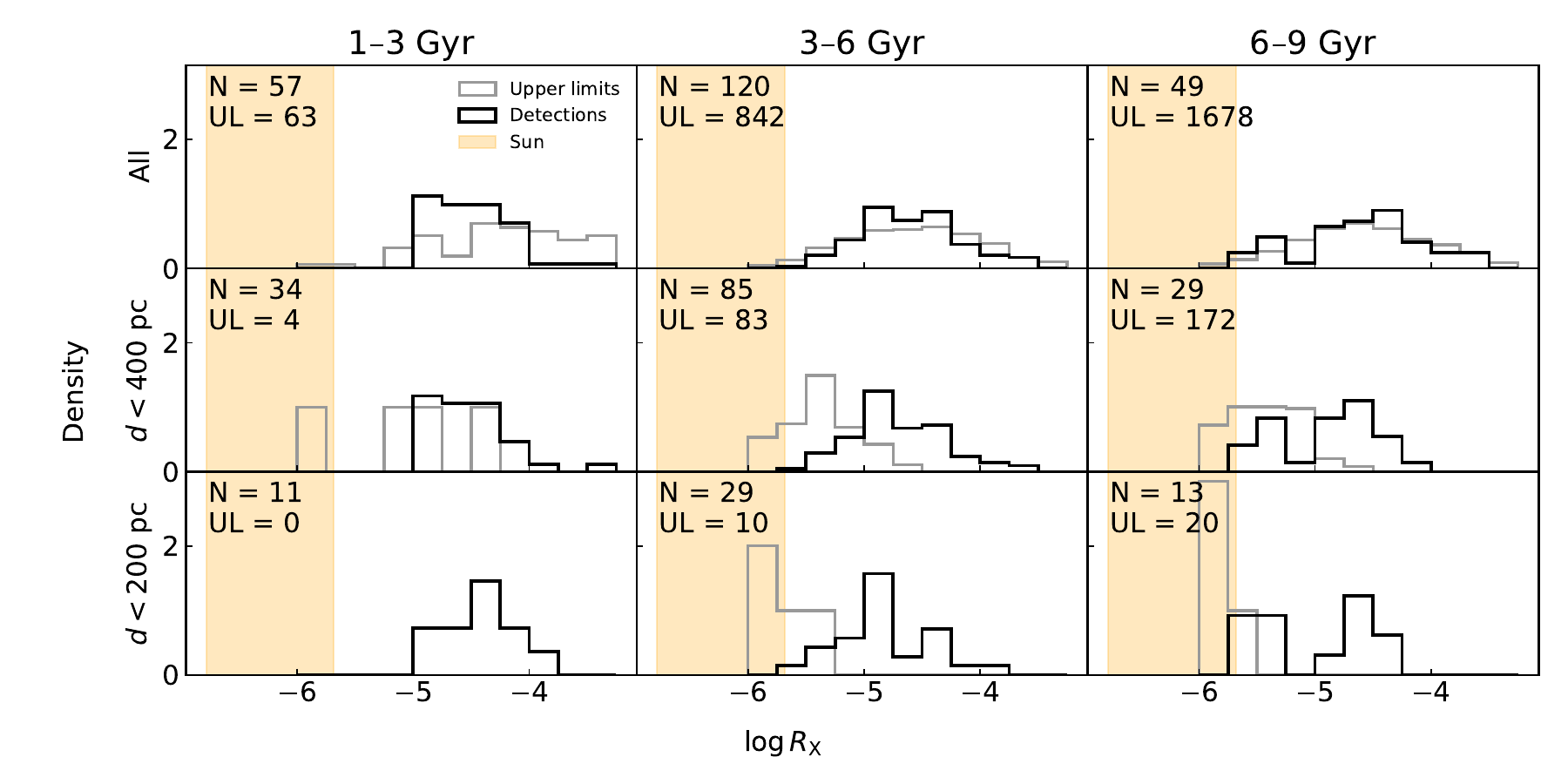}
    \caption{Histogram of $\log R_{\rm X}$ for solar analogs. From left to right, the panels correspond to age bins of 1–3 Gyr, 3–6 Gyr, and 6–9 Gyr. From top to bottom, the panels show the full solar-analog sample, the subsample within 400 pc, and the subsample within 200 pc, respectively. The black and gray histograms represent the distributions of detections and upper limits, respectively. The light orange shaded region indicates the range of solar $\log R_{\rm X}$ variability inferred from the observed X-ray luminosity of the Sun.}
    \label{logRx_hist_age_dist.fig}
\end{figure*}

We selected a subsample of solar analogs from the field-star sample, defined by $T_{\rm eff}=5680-5880$ K, $\log g = 4.4 \pm 0.3$ dex, and ${\rm [Fe/H]} = 0.0 \pm 0.3$ dex, yielding a total of 231 stars. These solar analogs were grouped into age bins, and the distributions of $\log R_{\rm X}$ were examined and compared with the solar value.
Previous studies have shown that the Sun is at the lower-activity end of the Ca~II~H\&K distribution for solar analogs, but remains consistent with the overall population at comparable ages \citep[e.g.,][]{2008ApJ...687.1264M,2020MNRAS.491..455B}. 
In the X-ray regime, the Sun lies near the low-activity end of the solar-analog distributions (Figure~\ref{logRx_hist_age_dist.fig}). Simultaneously, a group of solar analogs have upper limits close to the activity level of the Sun. This indicates that while the present-day solar corona is indeed X-ray faint compared to typical solar analogs \citep{2016ApJ...830...44N}, there are still many solar analogs that share similar activity levels.

\subsection{Multi-Wavelength Evolution of Stellar Activity}
\label{compare.sec}

\begin{figure*}[!htbp]
    \centering
    \subfigure[]{
    \includegraphics[height=0.9\textwidth]{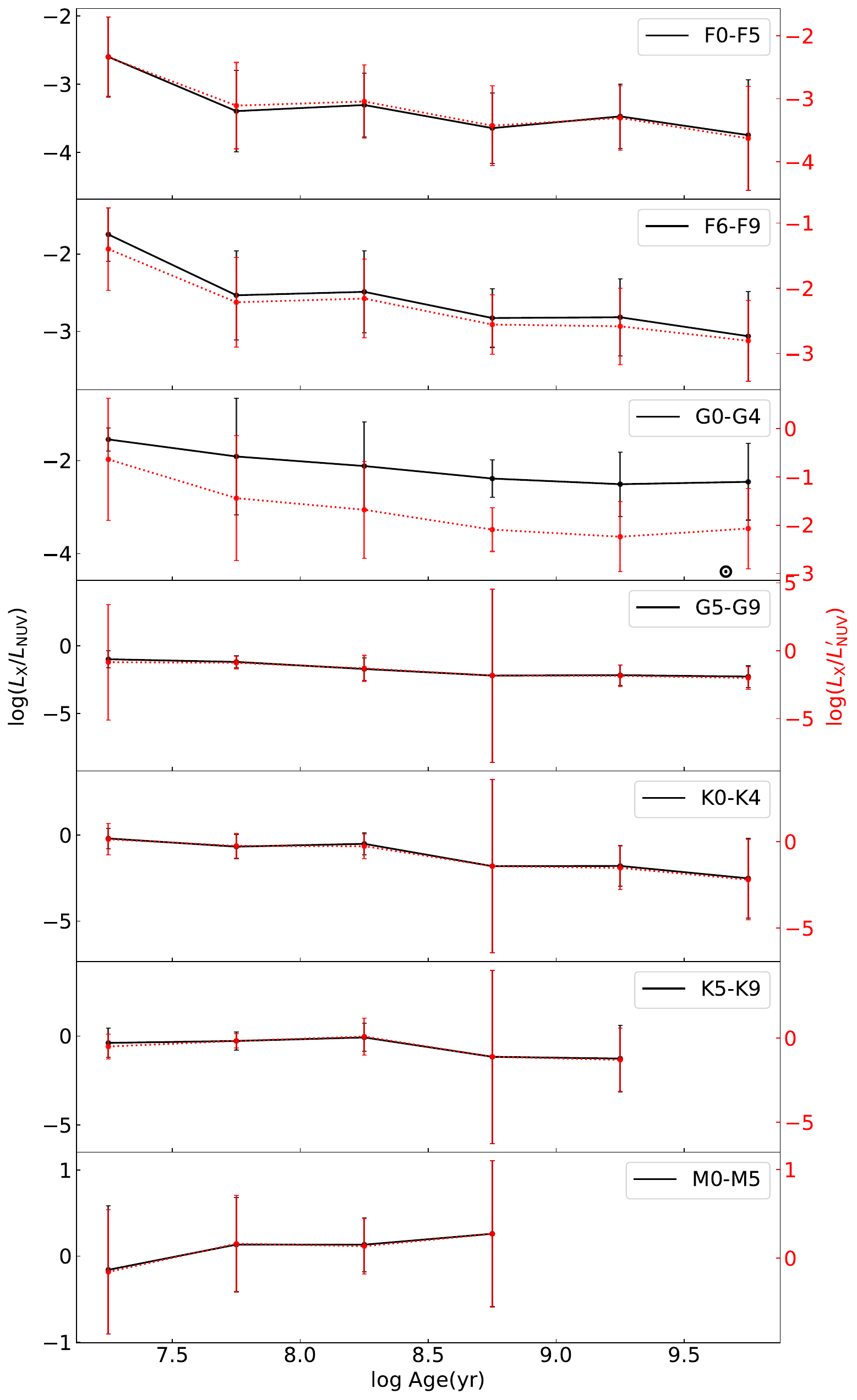}
    \label{lx_lnuv.fig}}
    \subfigure[]{
    \includegraphics[height=0.9\textwidth]{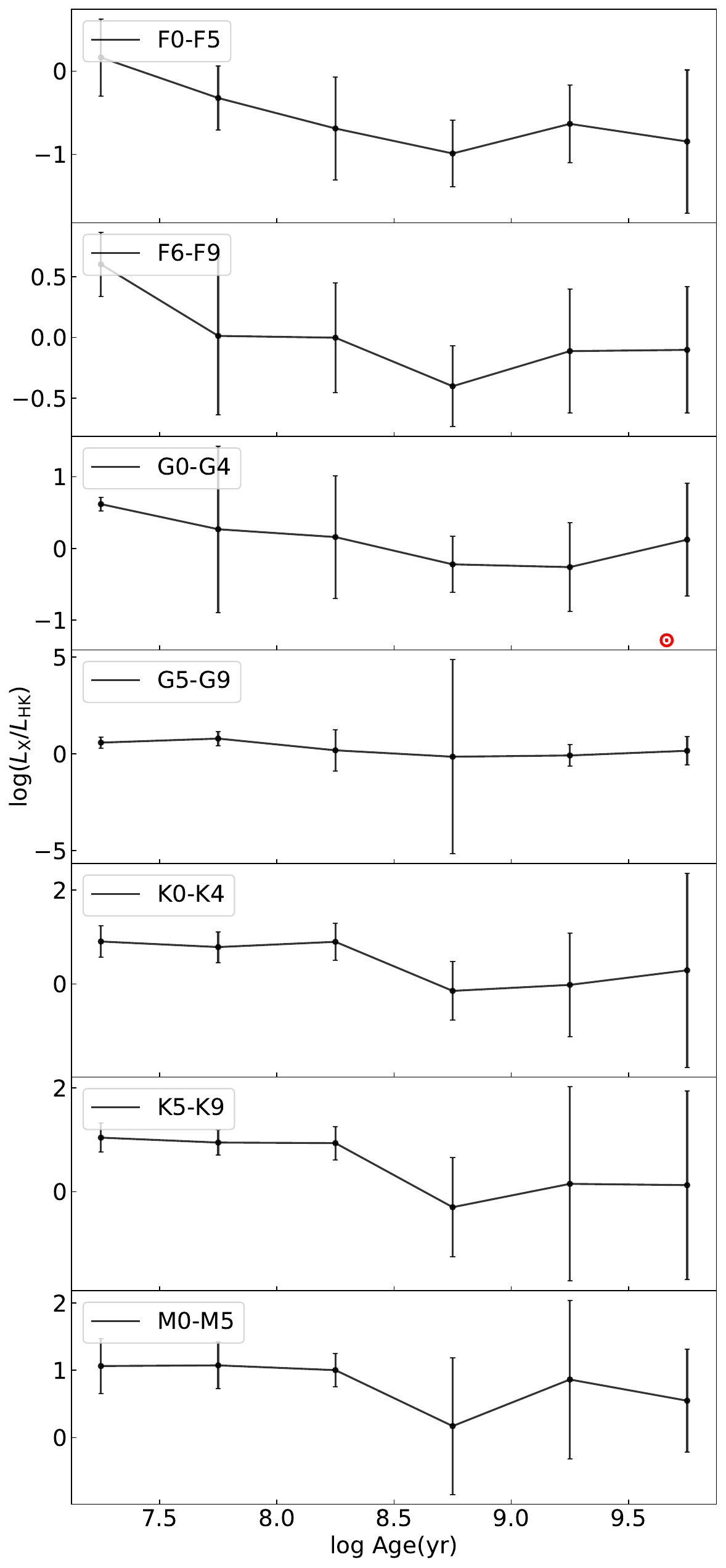}
    \label{lx_rhk.fig}}
    \caption{The evolution of $L_{\rm X}/L_{\rm NUV}$ (Panel a) and $L_{\rm X}/L_{\rm HK}$ (Panel b) for different stellar types.}
    \label{multi_evo.fig}
\end{figure*}

Stellar magnetic activity manifests across different atmospheric layers, from the photosphere to the hot corona, and is therefore the best characterized through a combination of diagnostics at different wavelengths. X-ray emission traces coronal plasma heated to several million Kelvin, ultraviolet emission originates predominantly from the chromosphere and transition region, and Ca~II~H\&K emission probes the lower chromosphere. Comparing activity indicators across these wavelength regimes allows us to investigate how magnetic energy is redistributed among different atmospheric layers as stars evolve. In Paper~I and Paper~II, we have investigated the evolution of stellar UV and Ca~II~H\&K emission. Here we compare of these different activity indicators. 

Using the stellar sample established in Paper~I, we investigated the multi-wavelength evolution of stellar activity by constructing age-binned median sequences for different spectral types. We divided the sample into logarithmic age bins spanning $\log{\rm Age}=6.5$ to 10, with a bin width of 0.5~dex, and computed the median value of near-ultraviolet (NUV) luminosity, $\log L_{\rm NUV}$, for each spectral type within each age bin.

To obtain a clearer NUV chromospheric activity indicator, we corrected the observed NUV emission for the photospheric contribution. Following the prescription of \citet{2024ApJ...976...43W}, we estimated the photospheric NUV luminosity, $L_{\rm NUV,ph}$, by matching the stellar atmospheric parameters provided by the StarHorse catalog \citep{2018MNRAS.476.2556Q}. We then computed the excess NUV emission, $\log L^{\prime }_{\rm NUV} = \log(L_{\rm NUV} - L_{\rm NUV,ph})$, and derived its median value for each spectral type in each age bin.

We also calculated the evolution of chromospheric activity traced by the Ca II H\&K lines. Using the Ca II H\&K activity indicators derived from LAMOST spectra (Paper~II), we similarly divided the sample into logarithmic age bins from $\log{\rm Age}=6.5$ to 10 with a step of 0.5~dex, and calculated the median $R'_{\rm HK}$ value for each spectral type. For the purpose of comparing coronal and chromospheric activity across wavelengths, we approximate the ratio $L_{\rm X}/L_{\rm HK}$ by the commonly used activity index ratio $R_{\rm X}/R'_{\rm HK}$, under the assumption that the bolometric normalization in the calculations is roughly equal. 

Figure~\ref{multi_evo.fig} presents the evolution of $L_{\rm X}/L_{\rm NUV}$, $L_{\rm X}/L^{\prime}{\rm NUV}$, and $L{\rm X}/L_{\rm HK}$ for different spectral types, which trace the relative efficiency of coronal heating compared to chromospheric emission. For early-type stars (F and early G), both $L_{\rm X}/L_{\rm NUV}$ and $L_{\rm X}/L_{\rm HK}$ show a possible decrease with age, though with relatively large uncertainties. Whereas later-type stars (late G to M) maintain approximately constant ratios over much longer timescales, suggesting a slower evolution of coronal activity relative to chromospheric indicators.

Furthermore, when comparing stars across the full F–M spectral range, both $L_{\rm X}/L_{\rm NUV}$ and $L_{\rm X}/L_{\rm HK}$ systematically increase toward later types: $\log L_{\rm X}/L_{\rm NUV}$ values increases from $-3$ for F-type stars to 0 for M-type stars, while $\log L_{\rm X}/L_{\rm HK}$  values increases from $-0.5$ for F-type stars to $1$ for M-type stars. This trend implies that a progressively larger fraction of magnetic energy is released through coronal emission in lower-mass stars. It may indicate that either a greater portion of magnetic energy reaches the corona, the efficiency of coronal heating is higher, or both effects contribute \citep{2011ApJ...743...48W}. A similar phenomenon has been noted in previous studies \citep[e.g.,][]{2014AJ....148...64S, 2023ApJ...951...44R}, consistent with the idea that coronal emission becomes increasingly dominant in late-type stars \citep{2004A&A...416..713G}.

\begin{figure*}
    \centering
    \subfigure[]{
    \includegraphics[width=0.48\textwidth]{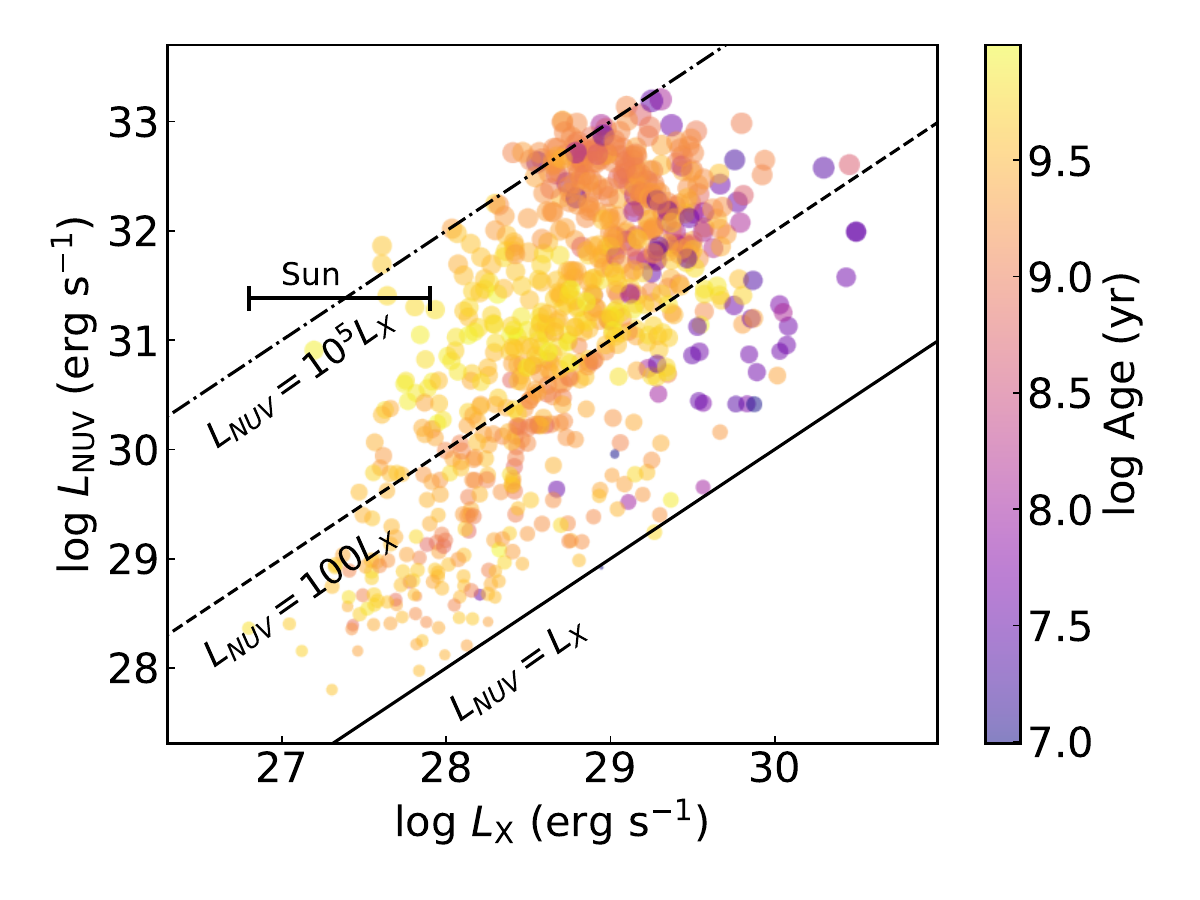}
    \label{lx_vs_lnuv.fig}}
    \subfigure[]{
    \includegraphics[width=0.48\textwidth]{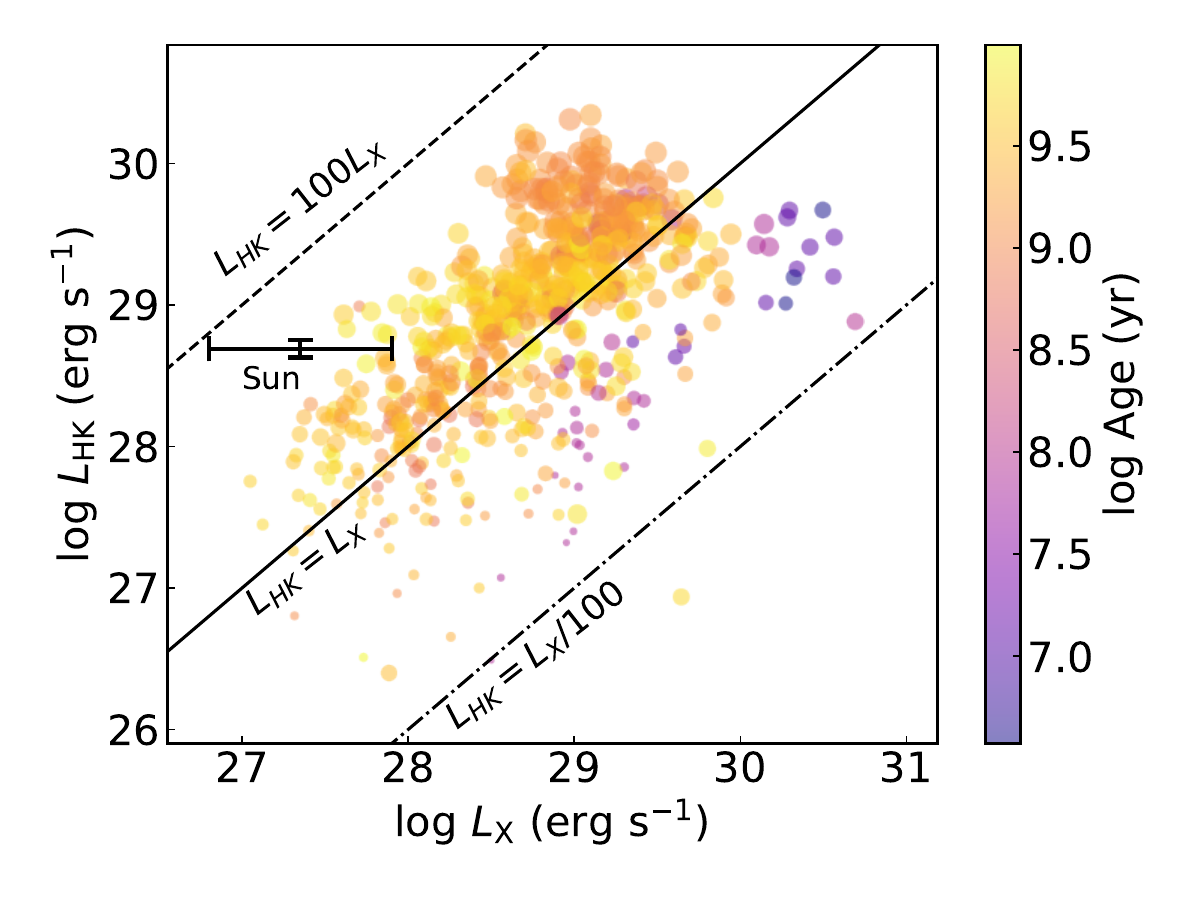}
    \label{lx_vs_lhk.fig}}
    \caption{Panel (a): Relation between $L_{\rm NUV}$ and $L_{\rm X}$. Points are color-coded by stellar age, and the symbol size scales with stellar effective temperature. The black bar marks the Sun \citep{2003ApJ...593..534J, 2008ApJ...687.1264M}. 
    Panel (b): Relation between Ca II H\&K luminosity $L_{\rm HK}$ and X-ray luminosity $L_{\rm X}$. }
    \label{compare.fig}
\end{figure*}

We also cross-matched the X-ray sample with the UV sample established in Paper~I and Ca~II~H\&K sample in Paper~II, yielding 823 and 713 common sources, respectively. For these matched samples, we examined the correlations between $L_{\rm NUV}$ and $L_{\rm X}$, as well as between $L_{\rm HK}$ and $L_{\rm X}$. For consistency, the Ca~II~H\&K luminosity was approximated as $L_{\rm HK} = R_{\rm HK}\times L_{\rm bol}$. The comparison results are shown in Figure~\ref{compare.fig}. Both $L_{\rm NUV}$ and $L_{\rm HK}$ exhibit clear positive correlations with $L_{\rm X}$, indicating that chromospheric, transition-region, and coronal emissions are closely linked. The $L_{\rm X}/L_{\rm NUV}$ and $L_{\rm X}/L_{\rm HK}$ values are the functions of stellar mass and age. Notably, despite the overall correlations, the Sun is located toward the lower $L_{\rm X}$ side of the distribution at given chromospheric luminosity, indicating relatively weaker coronal emission compared to other solar analogs.

\subsection{XUV Evolution and Implications for Planetary Habitability}
\label{habitability.sec}

High-energy stellar radiation (XUV, 0.1--92 nm) in the combined X-ray and extreme-ultraviolet bands plays a fundamental role in regulating the long-term atmospheric evolution and habitability of planets. XUV photons drive upper-atmospheric heating, ionization, and hydrodynamic escape, and are therefore a primary agent of atmospheric erosion, particularly during the early evolutionary stages of planetary systems \citep[e.g.,][]{2009A&A...506..399L,2021A&A...649A..96J}. 

We estimated stellar XUV luminosity ($L_{\rm XUV}$) by combining the measured X-ray luminosity calculated with empirical relations in \citet{2021A&A...649A..96J}. The EUV fluxes in two wavelength intervals were derived as
\begin{align}
\log F_{\rm EUV,1} &= 2.04 + 0.681\,\log F_{\rm X}, \\
\log F_{\rm EUV,2} &= -0.341 + 0.920\,\log F_{\rm EUV,1},
\end{align}
where $F_{\rm EUV,1}$ and $F_{\rm EUV,2}$ correspond to the $10-36$~nm and $36-92$~nm wavelength ranges, respectively. So the total XUV luminosity in this work was then computed as
\begin{equation}
L_{\rm XUV} = L_{\rm X} + L_{\rm EUV,1} + L_{\rm EUV,2}.
\end{equation}

We then applied MCMC fitting to characterize the age-dependent evolution of $L_{\rm XUV}$ across different spectral types, following the method in Section~\ref{evo.sec}. The fitting results of $\log L_{\rm XUV}$ are shown in Table~\ref{fits_results.tab}. 
Based on the resulting evolutionary relations, we calculated the cumulative XUV energy emitted by the star until the end of the main sequence,
\begin{equation}
E_{\rm XUV}(t)=\int L_{\rm XUV}(t)\,dt.
\end{equation}

\begin{figure}[t]
\centering
    \includegraphics[width=0.48\textwidth]{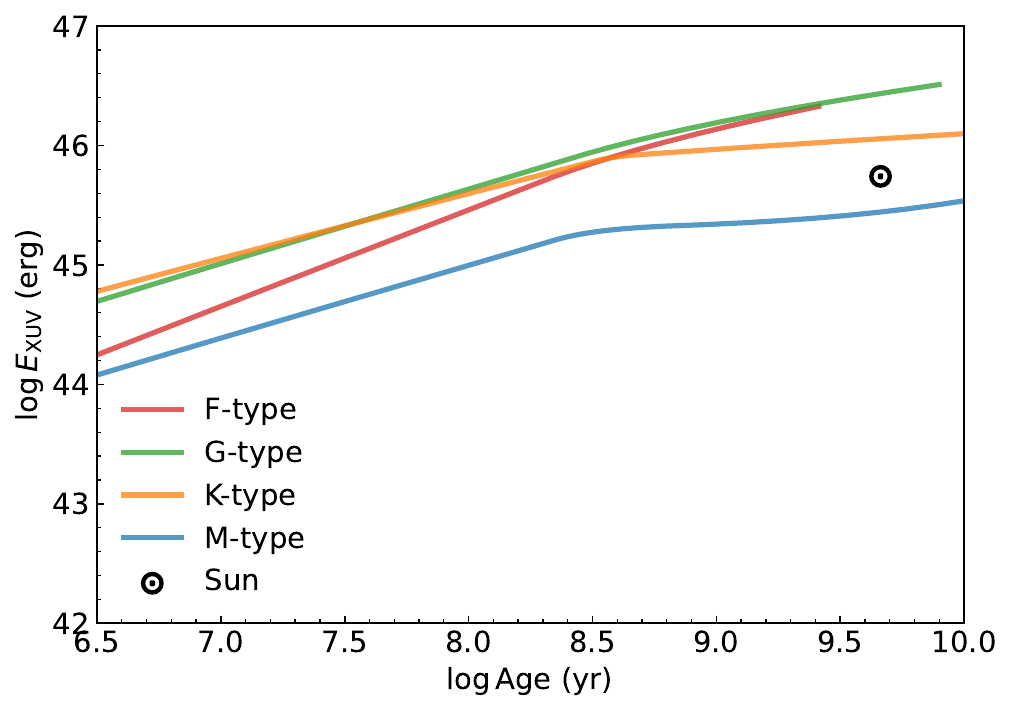}
    \caption{Time evolution of the cumulative XUV energy emitted by stars of different spectral types based on the fitted $L_{\rm XUV}$--age relations. The evolutionary phase was extrapolated back to $\log \rm Age (yr) = 6.5$ assuming an unchanged slope.}
    \label{exuv_evo.fig}
\end{figure}

The resulting evolution of the integrated XUV energy is presented in Figure~\ref{exuv_evo.fig}. Notably, the Sun lies below the median $E_{\rm XUV}$ evolution track of G-type stars, indicating that the present-day Sun is relatively XUV-quiet compared to typical solar-mass stars of similar age.

\begin{figure}[htbp]
\centering
    \includegraphics[width=0.46\textwidth]{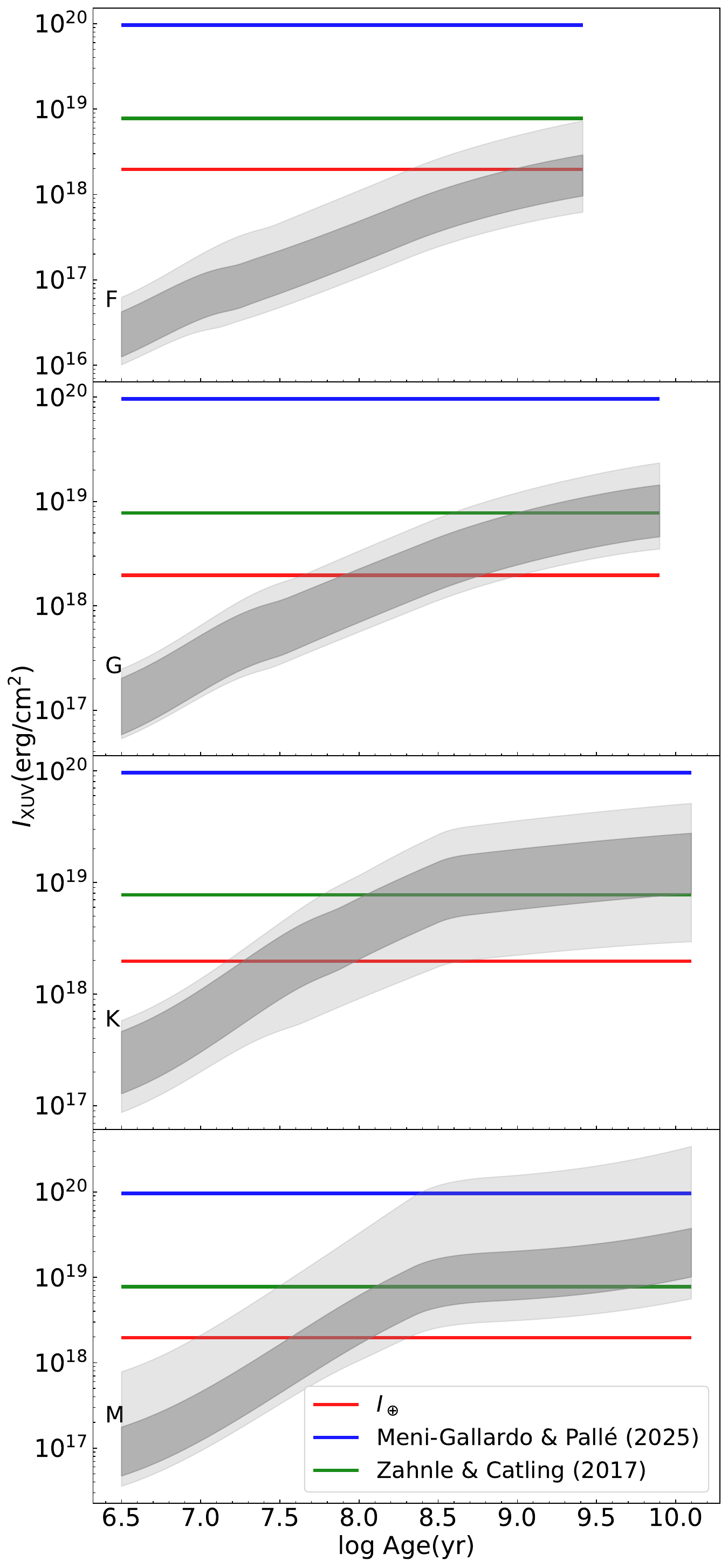}
    \caption{Time evolution of the cumulative XUV irradiation ($I_{\rm XUV}$) received by planets located within the CHZ for different stellar spectral types. The dark gray shaded region indicates the CHZ range derived using canonical stellar parameters of mid-subtype stars, while the light gray shaded regions represent the broader CHZ ranges obtained for earlier- and later-subtype stars. The red horizontal line is the cumulative XUV irradiation received by the Earth. The green and blue horizontal lines denotes the cosmic shoreline boundaries defined by \citet{2017ApJ...843..122Z} and \citet{2025arXiv250812865M}, respectively. The evolutionary phase was extrapolated back to $\log \rm Age (yr) = 6.5$ assuming an unchanged slope.}
    \label{xuv_evo.fig}
\end{figure}

To further assess the implications for planetary habitability, we combine the XUV evolutionary histories derived here with the evolution of circumstellar habitable zone (CHZ) obtained in Paper~I. We quantify the cumulative XUV irradiation received by a planet as
\begin{equation}
I_{\rm XUV} = \int \frac{L_{\rm XUV}(t)}{4\pi a^2} dt ,
\end{equation}
where $a$ is the star-planet distance corresponding to the inner and outer boundaries of the CHZ. This cumulative irradiation represents the total XUV energy per unit area on a planet over its evolutionary history and thus provides the long-term atmospheric irradiation experienced by planets in CHZ. The resulting cumulative $I_{\rm XUV}$ values integrated from 0.01~Myr to different stellar ages are presented in Figure~\ref{xuv_evo.fig}. 
For reference, we also plot the cumulative XUV irradiation received by the Earth over the age of the Solar System \citep{2017ApJ...843..122Z}. The solar XUV luminosity was estimated using the two-phase evolutionary prescription for G-type stars from \citet{2009A&A...506..399L}, and the corresponding cumulative XUV irradiation was then computed using the same equation described above.
We further marked the critical $I_{\rm XUV}$ threshold associated with the escape velocity of the Earth, derived from the empirical cosmic shoreline boundary. The cosmic shoreline defines a boundary in cumulative stellar irradiation versus planetary escape velocity, separating planets capable of retaining substantial atmospheres from those that undergo efficient atmospheric erosion \citep{2017ApJ...843..122Z,2025arXiv250812865M}. Comparing cumulative irradiation histories with this threshold therefore provides a direct assessment of whether planets within the CHZ are likely to preserve long-lived atmospheres.

Figure~\ref{xuv_evo.fig} shows the cumulative XUV irradiation of the planets at the CHZ boundaries. For F-type stars, planets residing within the CHZ consistently receive cumulative XUV irradiation lower than that experienced by the Earth, implying relatively weak atmospheric erosion over gigayear timescales. For G-type stars, CHZ planets reach Earth-like $I_{\rm XUV}$ levels after $\sim$100~Myr and always remain below the cosmic shoreline throughout their evolution, indicating favorable conditions for long-term atmospheric retention.

In contrast, a markedly different evolutionary picture emerges for K- and M-type stars. By the time these stars evolve the main-sequence phase, planets within their CHZs are exposed to a cumulative XUV that can exceed the cosmic shoreline threshold defined by \citet{2017ApJ...843..122Z}. Under this criterion, these planets may experience much stronger atmospheric escape and erosion, posing challenges for the long-term maintenance of Earth-like surface habitability. However, we note that the cosmic shoreline defined by \citet{2025arXiv250812865M} is significantly higher than that defined by \citet{2017ApJ...843..122Z}. The updated threshold implies that the CHZs around K- and M-type stars can also have cumulative XUV energies well below the cosmic shoreline. 

In addition, energetic stellar flares can provide episodic bursts of high-energy radiation, temporarily enhancing the XUV flux received by orbiting planets \citep[e.g.,][]{2018ApJ...867...71L,2021NatAs...5..298C}. Recent studies showed that such flare-driven radiation may become increasingly important for planets at larger orbital distances \citep[e.g.,][]{2025ApJ...985..100D,2025A&A...702A.112C}, where quiescent XUV emission is insufficient to fully deplete their atmosphere.
Nevertheless, more observational and theoretical studies are needed to make a robust conclusion about the effects of XUV emission on habitability.

\section{Summary}
\label{summary.sec}

In this work, we presented a comprehensive investigation of stellar X-ray and XUV activity evolution and its implications for planetary habitability. Building upon the stellar sample in Paper~I, we constructed a sample of 2,960 F--M dwarfs (from open clusters and the field stars) with \textit{Chandra} and \textit{XMM-Newton} detections, and 8,753 stars with upper-limit estimates.

We found that the X-ray activity evolution depends on stellar mass. 
For F- and G-type stars, the overall evolution broadly follows the traditional picture: an early saturated (or weakly declining) phase, followed by a modest decline. 
In contrast, K- and M-type stars exhibit a distinct three-phase evolution, containing an early saturated phase, an intermediate phase of rapid decline, and a final modest-decay phase. 
The overall amplitude of the evolution is similar across spectral types, with the X-ray activity index ($R_{\rm X}$) decreasing by 2--3 orders of magnitude from young to old ages. 
We also found that although the Sun lies near the low-activity end of the solar-analog distributions, there are still some solar analogs with upper-limit estimates sharing similar activity levels.

By combining X-ray activity with ultraviolet and the Ca~II~H\&K band, we found that the ratios of $L_{\rm X}/L_{\rm NUV}$ and $L_{\rm X}/L_{\rm HK}$ decline more rapidly with age in earlier-type stars, while later-type stars maintain nearly constant ratios over long timescales. Moreover, the $L_{\rm X}/L_{\rm NUV}$ and $L_{\rm X}/L_{\rm HK}$ values increase toward lower-mass stars, indicating an increasing fraction of the magnetic energy is radiated in the X-ray regime. This result supports the conclusion that coronal emission becomes progressively more dominant relative to chromospheric emission in cooler stars.

Using the fitted activity--age relations, we calculated the cumulative XUV irradiation received by planets throughout stellar evolution. 
We found significant differences across stellar types in the cumulative XUV irradiation. Planets located in the CHZs of F-type stars receive lower cumulative XUV irradiation than the Earth, while those around G-type stars reach Earth-like XUV irradiation levels after the first $\sim$100 Myr and remain below the empirical cosmic shoreline adopted from \citet{2017ApJ...843..122Z}. 
In contrast, the XUV irradiation received by planets in the CHZs of K- and M-type stars would exceed the cosmic shoreline shortly after reaching the main sequence, indicating efficient atmospheric escape and posing significant challenges for the long-term maintenance of Earth-like atmospheres. 
However, a different conclusion may be drawn using the new cosmic shoreline calibration by \citet{2025arXiv250812865M}. 
Therefore, more observational and theoretical studies are required to assess the long-term impact of XUV irradiation on planetary habitability.

\section*{acknowledgements}

We thank the anonymous referee for helpful comments and suggestions that have significantly improved the paper. 
This work has made use of data obtained from the \textit{Chandra} Source Catalog, and software provided by the Chandra X-ray Center (CXC) in the CIAO software package. This work is also based on observations obtained with \textit{XMM--Newton}, an ESA science mission with instruments and contributions directly funded by ESA Member States and NASA. We acknowledge use of Astropy, a community-developed core Python package for Astronomy (Astropy Collaboration, 2013).
This work was supported by National Natural Science Foundation of China (NSFC) under grant No. 12588202, Strategic Priority Program of the Chinese Academy of Sciences undergrant No. XDB4100000, the National Key R\&D Programme of China (grant No. 2025YFF0510603), NSFC under grant Nos. 12273057/12422303/11833002/12090042, Science Research Grants from the China Manned Space Project with No. CMS-CSST-2021-A08, and the Fundamental Research Funds for the Central Universities (grant Nos. 118900M122, E5EQ3301X2, and E4EQ3301X2).

\bibliographystyle{aasjournal}
\bibliography{bibtex}{}

\begin{appendix}

\section{Data Reduction about upper limits}

\subsection{Upper Limits Selection}
\label{upperlimit_select.sec}
\renewcommand{\thefigure}{A\arabic{figure}}
\setcounter{figure}{0}

\begin{figure}[!htbp]
    \centering
    \includegraphics[width=0.5\textwidth]{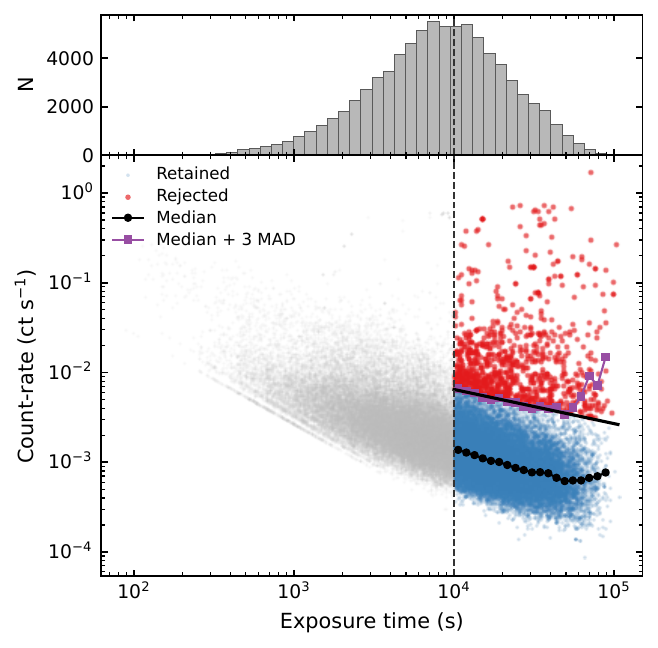}
    \caption{
    Upper limits versus exposure time. Gray points represent observations with exposure times shorter than 10 ks, which are excluded from the analysis. Black circles mark the median count rate in each exposure-time bin, while purple squares show the ``median$+3{\rm MAD}$". The black line is the linear fit to the purple squares. Red points lie above the fitted line and are considered anomalously high upper limits; blue points fall below the line and are retained for subsequent analysis.}
    \label{exposure_countrate.fig}
\end{figure}

Since a short exposure time may lead to an incorrect upper-limit estimate, we restricted our analysis to observations with exposure times longer than 10 ks. 
In addition, many of the upper limits are anomalously high, likely due to high background contamination. We therefore apply an empirical exclusion criterion. We grouped all upper limits into logarithmic exposure-time bins and, for each bin, calculated the median and the median absolute deviation (MAD). 
We then constructed an empirical $3\sigma$ upper envelope using the ``median$+3{\rm MAD}$" values, and fitted that as a function of exposure time (black line in Figure~\ref{exposure_countrate.fig}). Only measurements below the fitted envelope were retained for subsequent analysis. 

\subsection{Distance Completeness for the Upper Limit Sample}
\label{dist_distribution.sec}

\begin{figure}[!htbp]
    \centering
    \includegraphics[width=0.5\textwidth]{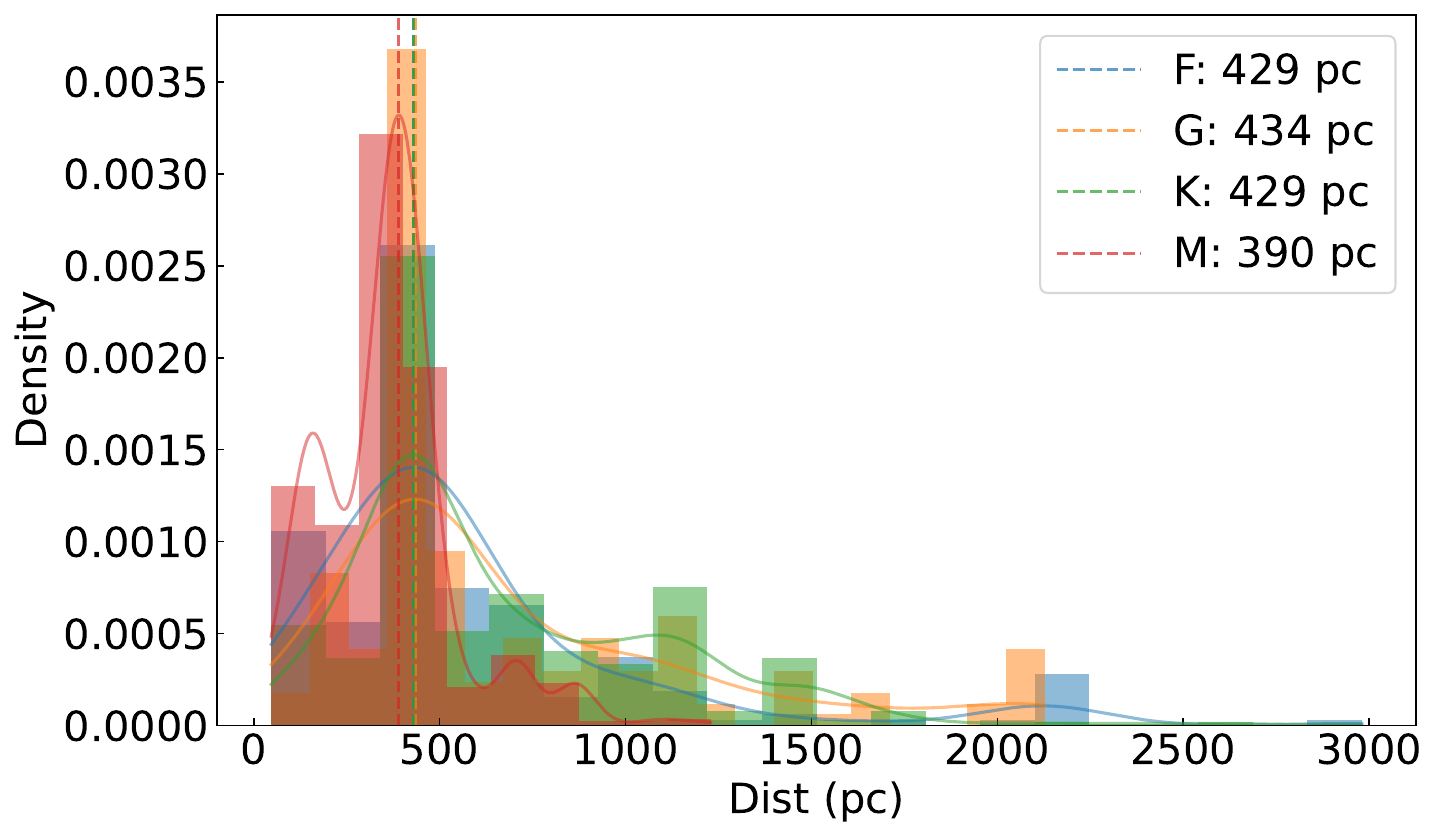}
    \caption{Distance distributions of the detected OC sample for F-, G-, K-, and M-type stars. The smooth curves show the corresponding KDE. Vertical dashed lines are the KDE peak distances. The vertical dashed lines indicate the adopted distance-completeness limits.}
\label{oc_dist_distribution.fig}
\end{figure}

To reduce distance-dependent sensitivity biases, we restricted the upper-limit sample using the distance-completeness limits inferred from the distance distribution of detected open-cluster sources. 
For each spectral type, we adopted the KDE peak distance of the detected OC sample as the completeness limit to select stars with X-ray upper limits.
The adopted limits are 429 pc for F-type stars, 434 pc for G-type stars, 429 pc for K-type stars, and 390 pc for M-type stars (Figure~\ref{oc_dist_distribution.fig}).

\subsection{The detection fraction for each spectral type}
\label{detected_frac.sec}

\begin{figure}[!hp]
    \centering
    \includegraphics[width=0.5\textwidth]{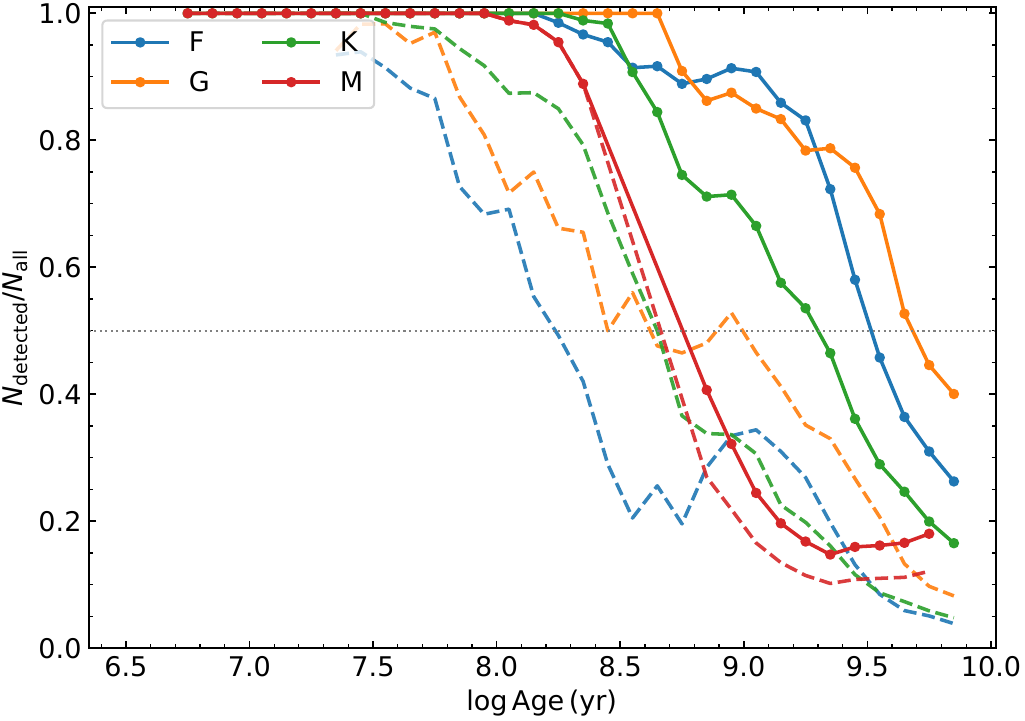}
    \caption{
    X-ray detection fraction as a function of age for different spectral types. The dashed lines show the full sample, while the solid lines show the sample after applying the distance-completeness limits. The horizontal dashed line marks $f_{\rm det}=0.5$.}
    \label{detected_to_all.fig}
\end{figure}

We examined how the detection fraction ($f_{\rm det}=N_{\rm det}/(N_{\rm det}+N_{\rm nondet})$) in each age bin varies with distance for each spectral type (Figure~\ref{detected_to_all.fig}). 
This reveals that the detection fraction is very low for old stars, indicating that a large fraction of the data are censored. In such cases, the Kaplan–Meier estimator may not robustly constrain the median values.

\clearpage

\section{Effect of Stellar Age Uncertainty on activity--age relation}
\label{field_age_error.sec}
\renewcommand{\thefigure}{B\arabic{figure}}
\setcounter{figure}{0}

\begin{figure}[!htbp]
    \centering
    \includegraphics[width=0.5\textwidth]{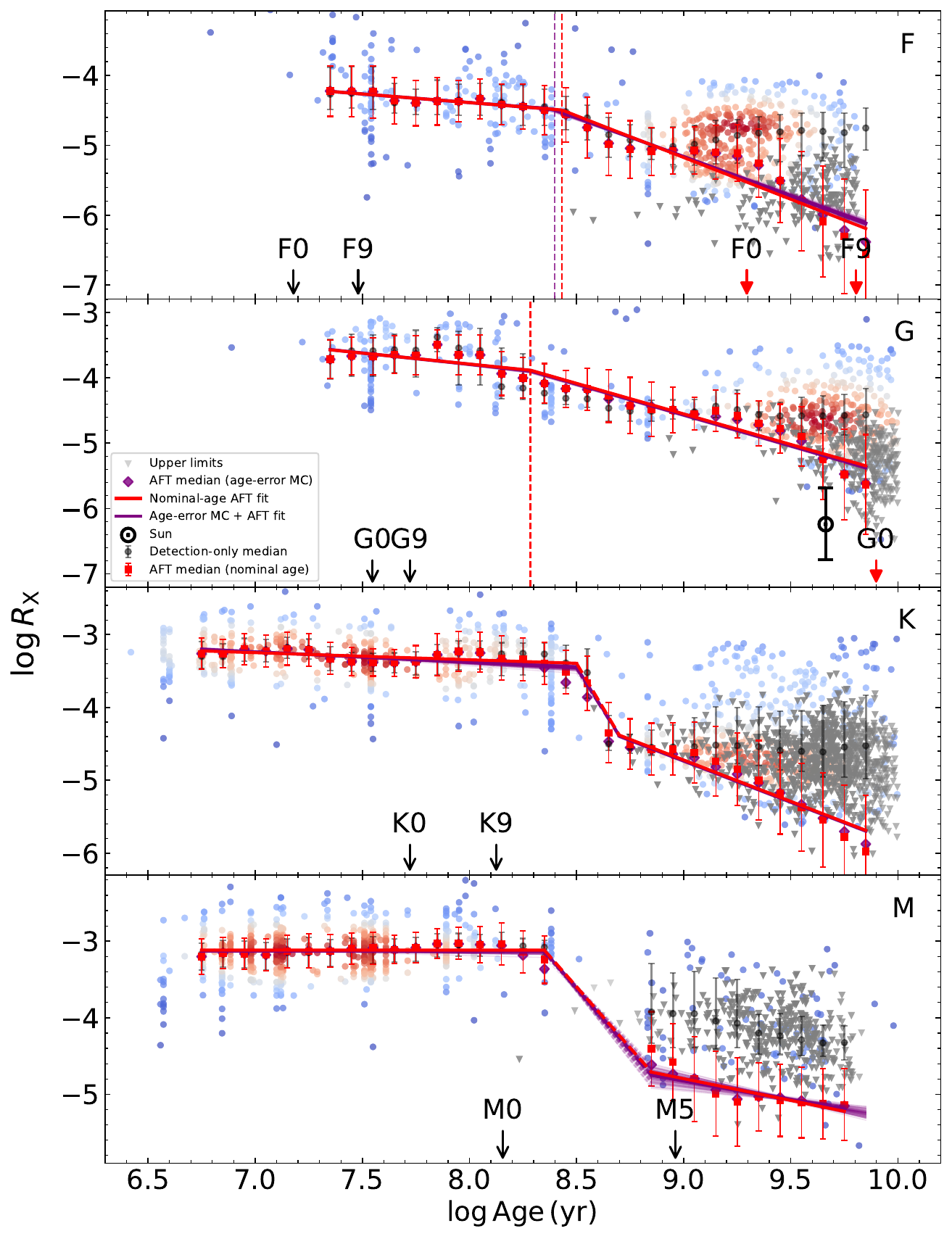}
    \caption{Comparison of the activity–age relations obtained using initial stellar ages and after considering their uncertainties. Red solid curves are initial-age AFT fits. 
    Purple translucent curves show fits obtained from the 100 age-error Monte Carlo realizations, while purple solid curves show fits to their median results.
    Other symbols are same as the Figure~\ref{distribution_xray_log_fit_mcmc.fig}.}
    \label{age_error_fit.fig}
\end{figure}

Figure~\ref{age_error_fit.fig} presents the Monte Carlo test described in Section~\ref{evo.sec}. 
We resampled the stellar ages by adding Gaussian noise according to the uncertainties of \citet{2025ApJS..280...13W}, refit the activity–age relation, and repeated this 100 times.

\end{appendix}

\end{CJK*}
\end{document}